\documentclass[preprint,aip,cha]{revtex4-1}

\usepackage[english]{babel}
\usepackage{float}
\usepackage{graphicx}
\usepackage{dcolumn}
\usepackage{bm}
\usepackage{amsmath}
\usepackage{amssymb}
\usepackage{physics}
\usepackage{datetime}
\usepackage{hyperref}
\usepackage{bigints}
\usepackage{siunitx}
\usepackage{subcaption}
\usepackage{caption}
 \usepackage{todonotes}
\usepackage{color, colortbl}
\usepackage{cleveref}
\usepackage{multirow}
\usepackage{diagbox}
\usepackage{rotating}
\usepackage[normalem]{ulem}
\definecolor{Gray}{gray}{0.9}

\begin{document}

\title{Azimuthal structures  and turbulent transport in Penning discharge}
\author{\firstname{M.}~\surname{Tyushev}}
\author{\firstname{M.}~\surname{Papahn Zadeh}}
\author{\firstname{V.}~\surname{Sharma}}
\author{\firstname{M.}~\surname{Sengupta}}
\affiliation{Department of Physics and Engineering Physics, University of Saskatchewan, Saskatoon SK S7N 5E2, Canada}
\author{\firstname{Y.}~\surname{Raitses}}\affiliation{Princeton Plasma Physics Laboratory, Princeton, New Jersey 08540, USA}
\author{\firstname{J.-P.}~\surname{Boeuf}}\affiliation{LAPLACE, Université de Toulouse, CNRS, INPT, UPS, 118 Route de Narbonne, 31062 Toulouse, France}
\author{\firstname{A.}~\surname{Smolyakov}}
\affiliation{Department of Physics and Engineering Physics, University of Saskatchewan, Saskatoon SK S7N 5E2, Canada}

\begin{abstract}
   Azimuthal structures  in cylindrical Penning discharge are studied with  2D3V radial-azimuthal PIC/MCC model with the axial magnetic field. The discharge  is  self-consistently supported by ionization due to the axial injection of  electrons.  It is shown that the steady-state  discharge can be supported in two different regimes with different type of observed azimuthal structures. The transition between the regimes is controlled by the mechanism of the  energy input to the discharge.
      In the first regime (low energy of the injected electrons), with the pronounced $m=1$ spoke activity,   the power input is dominated by the energy absorption due to the   radial current and self-consistent electric field. 
        In the other regime (higher energy of the injected electrons), with  prevalent small scale $m>1$ spiral structures, and the lower values of the anomalous transport,  the total energy deposited to the discharge  is lower and is  mostly due to the direct input of the kinetic energy from  the axial electron beam. 
       We show that the  large  (m=1) spoke  and small scale structures  occur as a result of Simon-Hoh and lower hybrid instabilities driven by the electric field, density gradient, and collisions.
    We show that the spoke frequency  follows the   equilibrium ion rotation frequency. 
   
\end{abstract}

\maketitle

 


\section{Introduction}

Magnetically enhanced plasma discharges are widely used in industry and various technologies. Regime of partially magnetized plasmas, when the electrons are magnetized and strongly confined by the magnetic field, while the ions are weakly magnetized and can be  controlled via the electric field offers many opportunities for various applications for high density plasmas. At the same time however, combination of the magnetic field and plasma gradients results  in
appearance of drift waves, instabilities, and turbulence. Plasma and energy transport across the magnetic field due to turbulent fluctuations and large scale structures are typically much larger than the classical values. Despite a wide usage of magnetically enhanced plasma sources in industry and long history of studies, the physics of fluctuations and anomalous transport in such devices  is still poorly understood impeding the further progress and development of new applications.

A plasma cylinder  with axial magnetic field is a prototype  configuration for the Penning discharges used in many technological  applications.  In this paper we study fluctuations and nonlinear structures in the cylindrical partially magnetized $\mathbf{E}\cross \mathbf{B}$ Penning plasma discharge using a 2D radial-azimuthal PIC-MCC  models.
The goal of the study is to clarify the mechanisms and characteristics of instabilities and transport in this system. Our study  emphasizes the role of self-consistent ionization as well as the physics of large (spokes) and small scale (spiral arms) azimuthal structures that appear in the nonlinear saturated state.  

We focus on the case when the discharge is self-consistently sustained by the energy input from the axially injected electrons representing the electron beam  from the RF cathode of the Penning discharge device \cite{RaitsesIEPC2015,RodriguezPoP2019}. We demonstrate that in  addition to the direct kinetic energy of the electron beam, the energy absorption due to the radial electron current and radial electric field  is an important mechanism  of the power input to the discharge. The total power delivered to the discharge and relative contributions of these two mechanisms define what type of azimuthal structures are excited  in the  discharge. 

 The potential in the discharge center is negative with respect to the walls due to the injection of beam electrons in this region. The potential drop depends on the current and energy of the beam electrons and is larger for lower beam energy because of the more efficient electron trapping by the magnetic field. Therefore, for lower beam electron energy, electron energy absorption is dominated by the radial current and self-consistent radial electric field. In these conditions, the simulations show that large scale structures, $m=1$ spokes are excited.  As the beam energy is increased, the potential drop is smaller, and the energy input becomes mostly due to the direct kinetic energy from the axial electron beam. In that case the small-scale spiral structures, with $m>1$, dominate.
We show that the  large and small scale structures  occur as a result of Simon-Hoh and lower hybrid type instabilities driven by the electric field, density gradient, and collisions \cite{SmolyakovPPCF2017}.  

In Section II, the simulations model and main parameters are described. In Section III, we investigate the role of collisions and ionization on the development and saturation of the self-consistent discharge and present the benchmark simulations with four different PIC codes.  The detailed study of characteristics of azimuthal structures and their relation to gradient-drift instabilities is presented in Section IV.  The parametric dependencies  of spoke  frequency on the magnetic field, box size, and ion species is studied in Section V. The regime with small scale spiral arms is demonstrated and analyzed in Section VI. Section VII presents the comparison of anomalous transport between two regimes. Summary and  discussion of the results are  provided in Section VIII.
 
\section{The simulation model}\label{sec_model}

\begin{figure}[htp]
\centering
\captionsetup{justification=raggedright,singlelinecheck=false}
\includegraphics[width=0.5\textwidth]{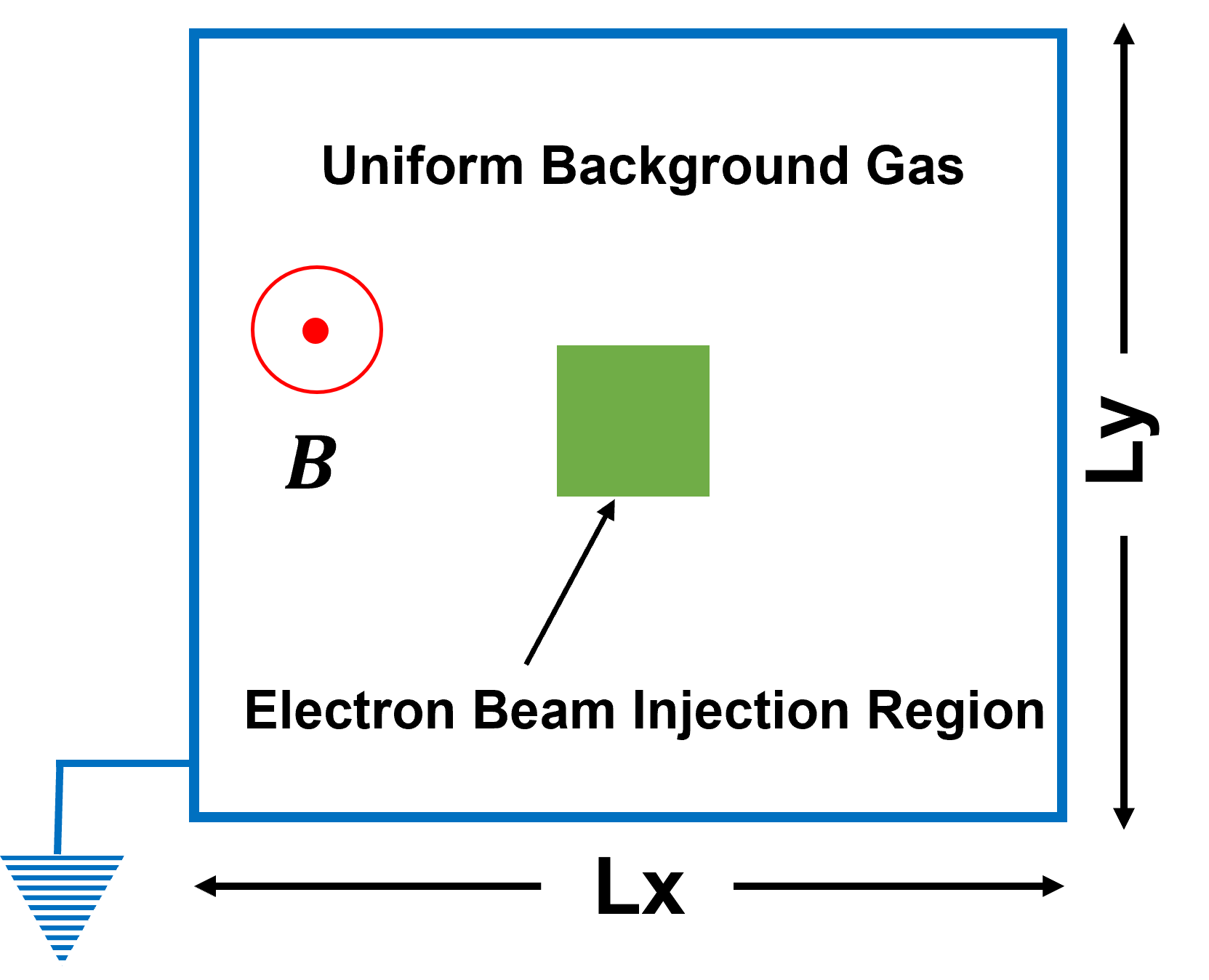}
\caption{The end view of the  cross-section of the modeled discharge.}
\label{fig:SimulationSetUp}
\end{figure}
In various applications with axial magnetic field both cylinder and rectangular cross-section configurations are used \cite{AbolmasovPSST2012,KimPSST2022,FubianiPoP2012,BoeufPoP2012a,BoeufPoP2012b,KimAIPAdv2021}. We employ  the rectangular geometry of the  discharge, as shown in \Cref{fig:SimulationSetUp}, as our base case.  We also  study the effect of the geometry and provide basic comparison of the result for  the rectangular, dodecagon, and cylindrical models in Appendix B.
 
The simulations are performed on a uniform Cartesian grid in the $x-y$ plane, with particles velocities in 3D $x-y-z$ space. A uniform axial magnetic field in the $z$ direction is applied perpendicular to the simulation domain with conducting  Dirichlet boundary condition for the potential, $\Phi=0$, and absorption for all incoming particles. 

A uniform electron beam is injected axially inside a square region in the center ($0.6\times0.6\;\text{cm}^2$). The density of the background neutral gas is uniform and constant throughout the simulations. The electron beam is represented by macro-particles introduced at every simulation time-step.  The beam electrons are injected with the same axial energy and zero temperature (zero energy spread) throughout the whole simulation. 

The simulations domain $L_x=L_y=3\;\text{cm}$  is split into $n_x \times n_y=250\times250$ grid cells in the $x-$ and $y-$directions. A typical injection region is a $50\times50$ cells square domain in the center of the box. This implies mesh resolution $\Delta x=0.12\;\text{mm}$ that resolves Debye length through the simulations which has been confirmed by direct calculation of electron temperature and plasma concentration in each cell.   

The simulation time step is $\Delta t=5\;\text{ps}$ that resolves electron cyclotron period and local plasma oscillation period by providing the practical criteria of $\omega_{pe,ce} \Delta t<0.5$. Electron cyclotron frequency is $\omega_{ce}=2.6\times10^9\;\text{s}^{-1}$, and the characteristic value of plasma frequency is $\omega_{pe}= 5.6\times10^9\;\text{s}^{-1}$ for  $n_p = 10^{16}\;\text{m}^{-3}$. The spatial grid and time step are set in way that the Courant-Friedrichs-Lewy (CFL) condition is satisfied with a large margin. For the fastest particles, which velocity is approximately equal to $2.4\times 10^6\;\text{m/s}$ we have: $CFL =  v \Delta t / \Delta x = 0.1 << 1$. The weight of a single macro-particle was equivalent to $1.7\times10^7$ $m^{-1}$  for simulations with Argon gas, and for convergence study with Hydrogen gas the number was $3\times10^6\, m^{-1}$, both for electrons and ions since on average plasma is quasineutral. This value results in low number of particles per cell (ppc) roughly equal to 3 (on average).  Further simulations in $6\times 6 \, cm^2$  geometry  were performed with a higher ppc number, demonstrating  good convergence as it is described below in Sections III,IV, and VI.

In simulations, the electrons are magnetized, with small Larmor radius $\rho_{e}<<L_x$, while ions remain  unmagnetized (with Larmor radius much larger  than plasma dimension $\rho_{i}>>L_x$).
We have tested the case of $3\times3\;\text{cm}^{2}$ box with magnetic field included in equations of motion for ions. Comparison with the case of unmagnetized ions showed that the effect is insignificant. Therefore, for simulations with the $6\times6\;\text{cm}^{2}$ box, magnetic field for ions was omitted. 

The main plasma parameters for our base case are given in the \Cref{table:PhysicParamBaseCas}. This is the regime when the large scale $m=1$ spoke structures are excited.  The spoke regime is studied in Sections III, IV, and V.  It is important to note that in this case, the energy of injected electrons is low compared to ionization energy, and the ionization occurs mainly through electrons heated by the power deposited in the radial electron current and electric field via the axial electron current. In the regimes with with higher energy of injected electron, when the ionization occurs directly from the injected electrons, small scales $m>1$ spiral arm structures dominate as discussed in Section VI. 

\begin{table}[htp]
\caption{Physical parameters for the base case (spoke regime) simulations with Argon.}
\begin{tabular}{|p{4cm}| p{2cm}| p{3cm}|}
\hline\hline 
\rowcolor{Gray}
\textbf{Property} & \textbf{Symbol} & \textbf{Value}  \\ \hline\hline
Magnetic Field & $B(\text{T})$ & $0.015$ \\\hline
Electron beam energy & $E_{b}(\text{eV})$ & $1.84$ \\ \hline
Electron beam current & $I_e(\text{A/m})$ &  $0.544$ \\ \hline
Neutral temperature & $T_n(\text{K})$ & $300$   \\ \hline
Neutral pressure & $P_n(\text{mTorr})$ & $40$ \\  \hline
Neutral density & $n_n(\text{m}^{-3})$ & $1.2875\times 10^{21}$  \\ \hline \hline
\end{tabular}
\label{table:PhysicParamBaseCas}
\end{table}

Our goal is to have the stationary discharge that is self-consistently maintained by  the electron beam.  As expected,  ionization and collisions significantly affect the discharge characteristics, such as plasma density and fluctuations energy in saturated steady-state.  We consider and compare the discharge evolution and  saturation for two scenarios: the case when only ionizing electron neutral collisions, and the case when the ionization, elastic  and excitation processes in electron neutral collisions are all included. 
The role of ionization and electron neutral collision was studied for two different neutral gases, Argon and Hydrogen.  The details of these investigations are presented in the following section.

In this study, we omit the ion-neutral collisions.
The purpose of this work is to describe, using a 2D PIC-MCC kinetic model,  the different types instabilities likely to develop in a Penning-type source, and to interpret them using theoretical dispersion relations. For this purpose, the physics of the problem has been simplified along these lines: 1) All ions created by ionization are assumed to reach the grounded electrode moving radially. In a real Penning source a fraction of the ions created by ionization are lost in the axial direction (typically on the  cathode side) along the z axis of our simulation domain. Since we do not simulate  the z-direction in our simulations, such axial losses are not included. Assumption that all the ions are collected radially by the grounded electrode would be a good approximation for a very long (with respect to Lx and Ly) discharge in the z direction. In principle, the axial losses could be simulated in 2D geometry by removing ions with with some prescribed rate, but it was not done in present manuscript.  2) We assume collisionless ions. This is not realistic for a 40 mtorr discharge, but allows a direct comparison with  results with dispersion relations with collisionless ions.  We have actually found that inclusion of ion collisions leads to more complex discharge regimes and instabilities and  higher  plasma density  in the center if axial ion losses are not included. The resulting increase of plasma density goes  beyond the values practical for available simulation resources. 
In this preliminary study we have neglected the axial losses and ion collisions and leave the study of these effects (and methods to represent in a 2D model the boundary conditions along the z axis) to a further work. 

\section{The role of ionization and collisions for the discharge self-sustainment}\label{BenchmarkDescription}

One of the goals  of this study was to study the large scale spoke instability in the discharge with self-consistent ionization. It was found that parameters  of the discharge, in particular, plasma density and fluctuation energy at  saturation  are rather sensitive to collision effects. We have employed several independent PIC-MCC codes to investigate the sources of the sensitivity and raise the confidence in the results  of the simulations. Descriptions of the codes used in this study are given in Appendix A. Though all codes follow general PIC-MCC methodology \cite{birdsall1991particle} there are differences in Poisson solvers and the way particle collisions are implemented. This convergence study reveals that although all codes demonstrate qualitatively similar results, there are some quantitative differences. In particular, we find that the implementation of the collision algorithm is a source of some differences. 
 
For the convergence studies, two  parameters: the total number of physical particles and electrostatic energy, were used as a metric for the comparison between four codes. We use prevalent electron-neutral  collisions: ionization, elastic, and excitation. Two cases were compared: the case  with only ionizing  collisions, and with ionizing and non-ionizing electron-neutrals collisions. As it was noted above, the ion-neutral collisions were omitted. 

For our base case, we consider the situation when the axial energy of injected electrons is below the ionization threshold. We show that the electrons gain energy through the work of the radial electric current, so they are heated, initiate the discharge, and support ionization. The axial electron current is maintained constant. The stationary values of plasma density and radial electric field in the discharge is established  self-consistently by the balance of the input energy against ionization and losses through the boundaries.  We inject electrons  with equal energy and zero thermal velocity. In the regime when the most power to the discharge is delivered by the electric field, it is even possible to maintain the discharge with injection of cold electrons. As long as the injected current remains the same, the characteristics of the discharge do not change much.  The convergence of the discharge to the saturated state was confirmed with all four codes. As it will be discussed below in Section VI, the transition to different regime occurs when the  injected current is reduced while the beam kinetic energy is increased, so the ionization is maintained directly by the energetic electrons of the beam rather than by the electrons heated in the discharge. 

\begin{figure}[htp]
\centering
\captionsetup[subfigure]{labelformat=empty}
\subcaptionbox{\label{fig:Benc_Num_Ionionly}}{\includegraphics[width=0.49\textwidth]{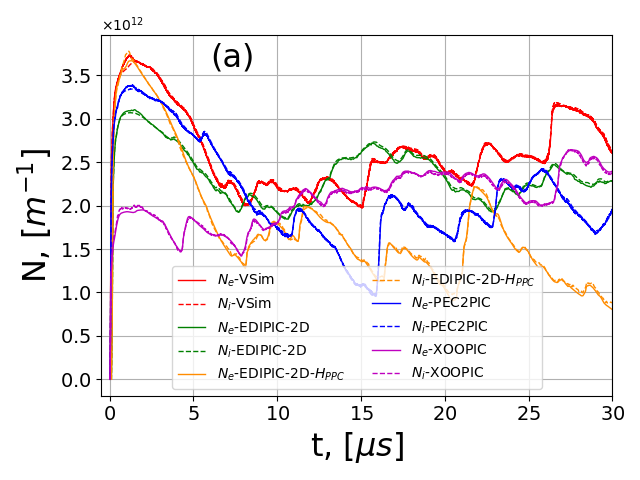}}
\subcaptionbox{\label{fig:Benc_Num_NoIon}}{\includegraphics[width=0.49\linewidth]{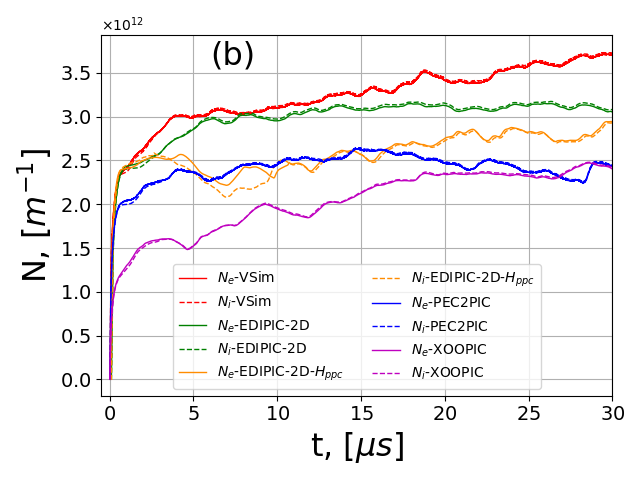}}
\captionsetup{justification=raggedright,singlelinecheck=false}
\caption{Total particles inventory as a function of time: a) case with only ionizing collision, b) case with ionization and non-ionizing electron neutral collisions. The results from different codes are shown by different colors; solid lines show electrons and dashed lines -- ions. The data with a $H_{ppc}$ label are from the   simulations with a larger larger number of computational particles per cell (ppc), which is around 85 for case (a) and 127 for the case (b).}
\end{figure}

\Cref{fig:Benc_Num_Ionionly} shows the time evolution of total electron and ion inventories in Argon discharge simulated with only ionizing collisions. All but XOOPIC codes show typical overshoot at the initial stage. The overshot recovers roughly by  $t\simeq 10\;\mu \text{s}$,  and all codes reach a steady state (after around $10\;\mu\text{s}$). The number of electrons and ions are close to each other, confirming quasi-neutrality in the  simulations, except early stage between $t=0.1\;\mu\text{s}$ to $t=0.5\;\mu\text{s}$ when the number of ions exceeds the number of electrons, due to electrons reaching the boundaries and being lost faster. In these simulations, the observed difference between the results of VSim, PEC2PIC, XOOPIC and EDIPIC-2D was about $10\%$ or less.

\begin{figure}[htp]
\centering
\captionsetup[subfigure]{labelformat=empty}
\subcaptionbox{\label{fig:BencHy_NUM_NoIon}}{\includegraphics[width=0.49\linewidth]{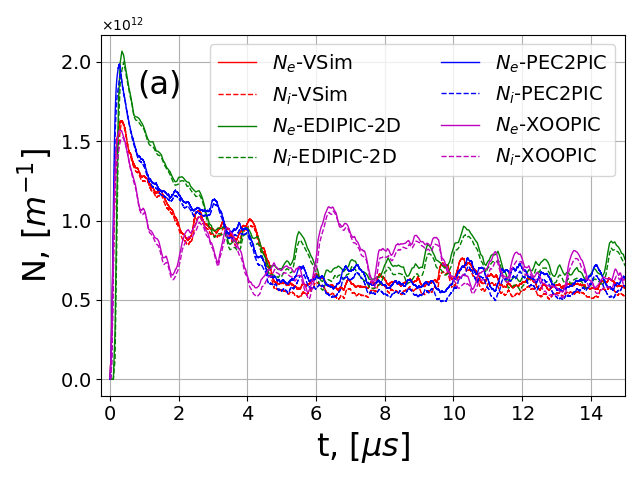}}
\subcaptionbox{\label{fig:BencHy_ESE_NoIon}}{\includegraphics[width=0.49\linewidth]{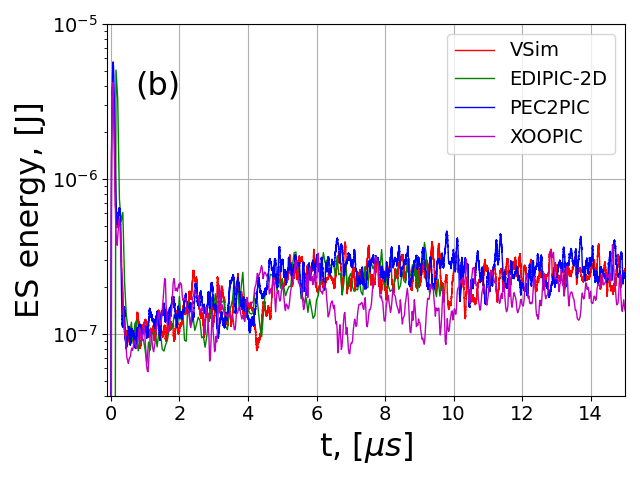}}
\captionsetup{justification=raggedright,singlelinecheck=false}
\caption{Temporal evolution of a) total particles inventory, b) electrostatic energy of Hydrogen gas for simulation with ionization plus non-ionizing electron neutrals collision. }
\end{figure}

\Cref{fig:Benc_Num_NoIon} shows the time evolution of particles inventory in the Argon discharge simulations with ionizing and non-ionizing electron neutral collisions included. In this case,  the number of particles  increases initially and reach steady state for all simulations codes (except VSim)  around $10\;\mu \text{s}$. The discrepancy between the results of VSim with EDIPIC and EDIPIC with PEC2PIC at $t=30\;\mu\text{s}$ in \Cref{fig:Benc_Num_NoIon} is almost $17\%$.  The behavior of plasma density in VSim and XOOPIC codes suggests that the saturation in these runs is slower and will be reached at later times. To confirm this, we have performed simulations for Hydrogen. As it is shown in \Cref{fig:BencHy_NUM_NoIon,fig:BencHy_ESE_NoIon},  the Hydrogen runs show good convergence for total number of particles and electrostatic energy of fluctuations for all codes.

\begin{figure}[htp]
\centering
\captionsetup[subfigure]{labelformat=empty}
\subcaptionbox{\label{fig:Benc_ESE_Ionionly}}{\includegraphics[width=0.49\linewidth]{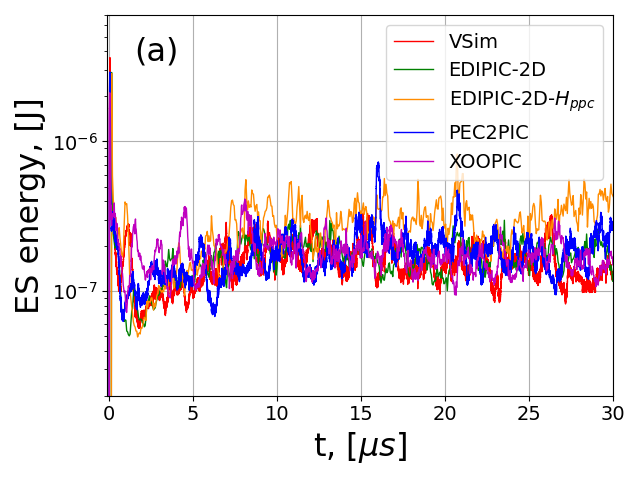}}
\subcaptionbox{\label{fig:Benc_ESE_NoIon}}{\includegraphics[width=0.49\linewidth]{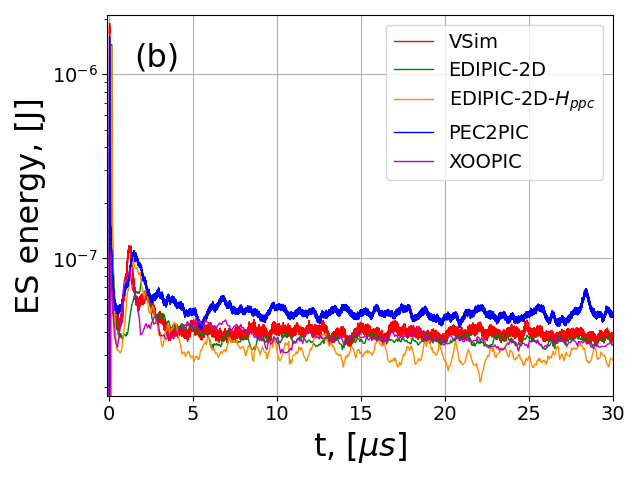}}
\captionsetup{justification=raggedright,singlelinecheck=false}
\caption{Evolution of electrostatic energy with time; a)  case  with only ionization collision, b) case with ionization plus non-ionizing electron-neutral collisions. The data with a $H_{ppc}$ label are from the simulations with a larger larger number of computational particles per cell (ppc), which is around 85 for the case (a), and 127 for the case (b).}
\label{BencESE}
\end{figure}

The evolution of the electrostatic energy (ES energy) is shown in \Cref{fig:Benc_ESE_Ionionly,fig:Benc_ESE_NoIon} for both  cases: with ionization only and with ionizing and non ionizing collisions.
One can see that  the saturation level of ES energy in simulation with only ionization collision is higher than for the case with ionization and non-ionizing electron neutral collisions.



\begin{figure}[htp]
\centering
\captionsetup[subfigure]{labelformat=empty}
\subcaptionbox{\label{fig:BencFUN_NUM_Ionionly}}{\includegraphics[width=0.49\linewidth]{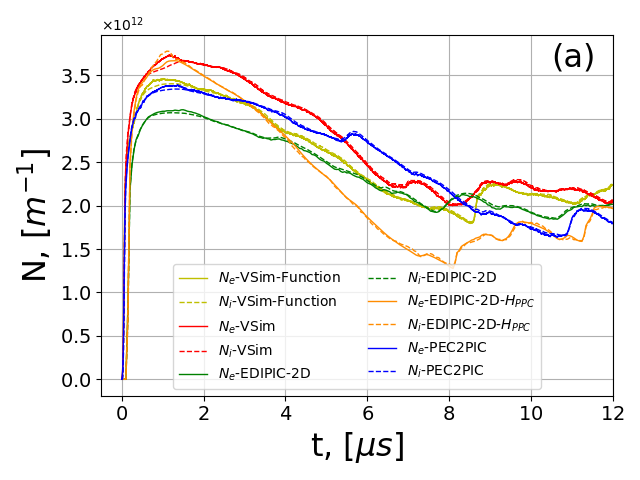}}
\subcaptionbox{\label{fig:BencFUN_NUM_NoIon}}{\includegraphics[width=0.49\linewidth]{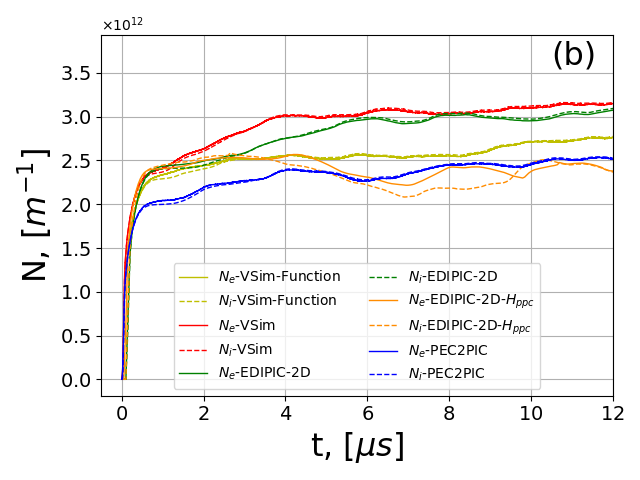}}
\captionsetup{justification=raggedright,singlelinecheck=false}
\caption{The effect of different implementation of the scattering cross-sections:  a)  Argon discharge with only ionization only; b) Argon discharge  with ionization plus non-ionizing electron neutral collisions. The yellow lines show the result from  VSim, using the same fit function as PEC2PIC. The data with a $H_{ppc}$ label are from the simulations with a larger larger number of computational particles per cell (ppc), which is around 85 for the case (a), and 127 for the case (b).}
\end{figure}

Discrepancies in results of presented simulations are  believed to be due to different ways the scattering cross section data are imported and interpolated. We performed two simulations in VSim to compare the results of two ways the cross sections are evaluated: one where the scattering cross-sections are imported from  an open-access database LXcat  and when the fitting function  for the cross sections is used.
The yellow lines in both \Cref{fig:BencFUN_NUM_Ionionly,fig:BencFUN_NUM_NoIon} show the evolution of particles inventory  for the VSim simulations with the cross section fitting function. The results show a good agreement between VSim and PEC2PIC when both use the same  MCC-cross section fit. Relevant details of the  cross-sections data and references are given in Appendix A.

To further confirm that different methods of importing the cross-section data lead to some discrepancies, we have conducted a experiment with Hydrogen as the neutral gas and used the linear interpolation between two points. The result presented in \Cref{fig:BencHy_NUM_NoIon,fig:BencHy_ESE_NoIon} show that all four codes give very similar results in the saturated state. The discrepancies at earlier times ($< 3\;\text{ms}$) are explained based on different behaviour of Poisson  solvers for un-resolved Debye length in the center and the artificial heating at this stage. At later stages, the density drops and the Debye length is well resolved across the whole simulation box.

\begin{figure}[htp]
\centering
\includegraphics[width=1.0\linewidth]{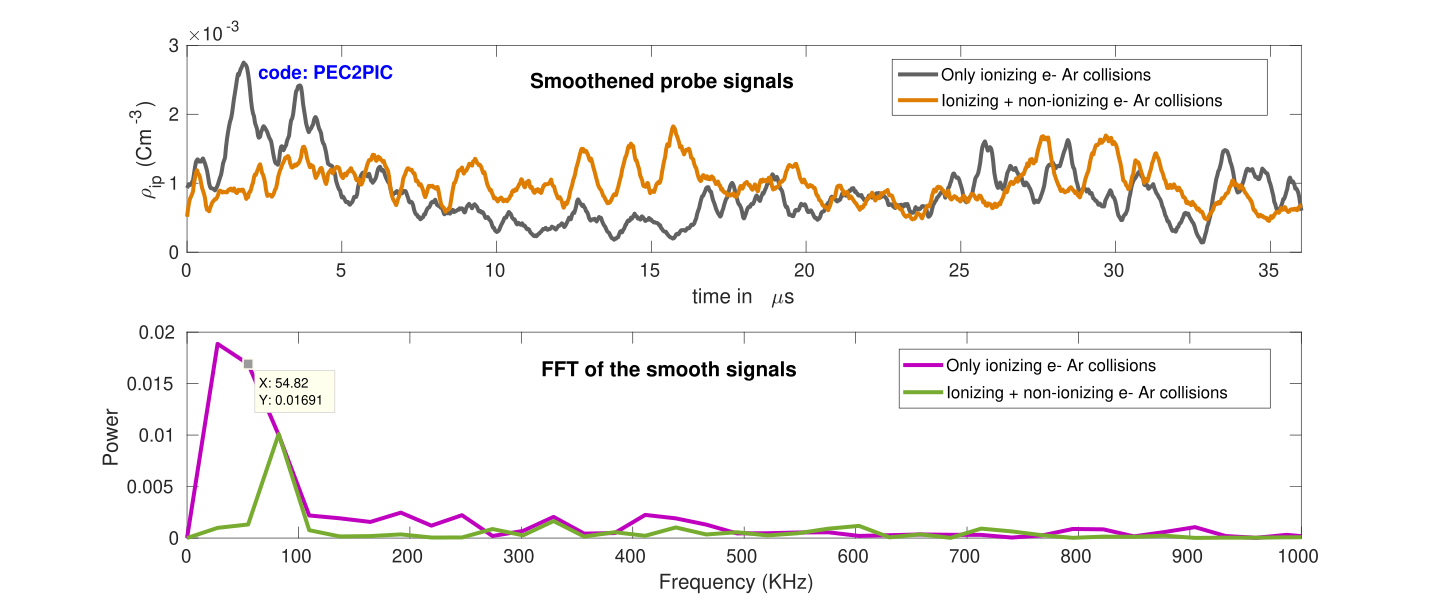}
\captionsetup{justification=raggedright,singlelinecheck=false}
\caption{The probe measurements results for  the ion charge  density,  $\rho_{ip}$, and power spectrum showing the effects of collisions on the spoke frequency.}
\label{fig:PEC2PICProbAnalysis}
\end{figure}

To investigate the role of ionization and collisions on the mode frequencies, we show in \Cref{fig:PEC2PICProbAnalysis} rotation frequency of the spoke from PEC2PIC simulations measured using Fast-Fourier-Transform (FFT) of a density probe's signal. The density probe measures local fluctuations in ion and electron densities at a location $0.084\;\text{cm}$ inside the left wall i.e. a distance of $0.028 L_x$ from the left wall, and a distance of $1.44\;\text{cm}$ from the bottom wall i.e. a height of $0.48 L_y$ referencing \Cref{fig:SimulationSetUp}. The probe picks up the spoke's front motion of the  $m=1$ mode as it rotates in the device's cross-section. Specifically two cases are shown: a) the Argon discharge simulated with only ionizing collisions b) the Argon discharge simulated with ionizing and non-ionizing electron neutral collisions i.e. elastic and excitation collisions. \Cref{fig:PEC2PICProbAnalysis}-top shows a smoothened version of the ion density fluctuations at the probe for the two cases while \Cref{fig:PEC2PICProbAnalysis}-bottom contains corresponding FFT's in time measuring the rotation frequencies. We get a rotation frequency of about $55\;\text{kHz}$ for PEC2PIC and $63\;\text{kHz}$ for EDIPIC-2D for the case with only ionization and about $82\;\text{kHz}$ and $77\;\text{kHz}$ correspondingly for the case with ionization plus non-ionizing collisions. The results show that the elastic electron scattering  increases the rotation frequency of the spoke. Furthermore, the case with only ionization has a wider frequency peak compared to the case with ionization plus non-ionizing electron neutral collisions. 



\begin{figure}[htp]
\centering
\includegraphics[width=1\linewidth]{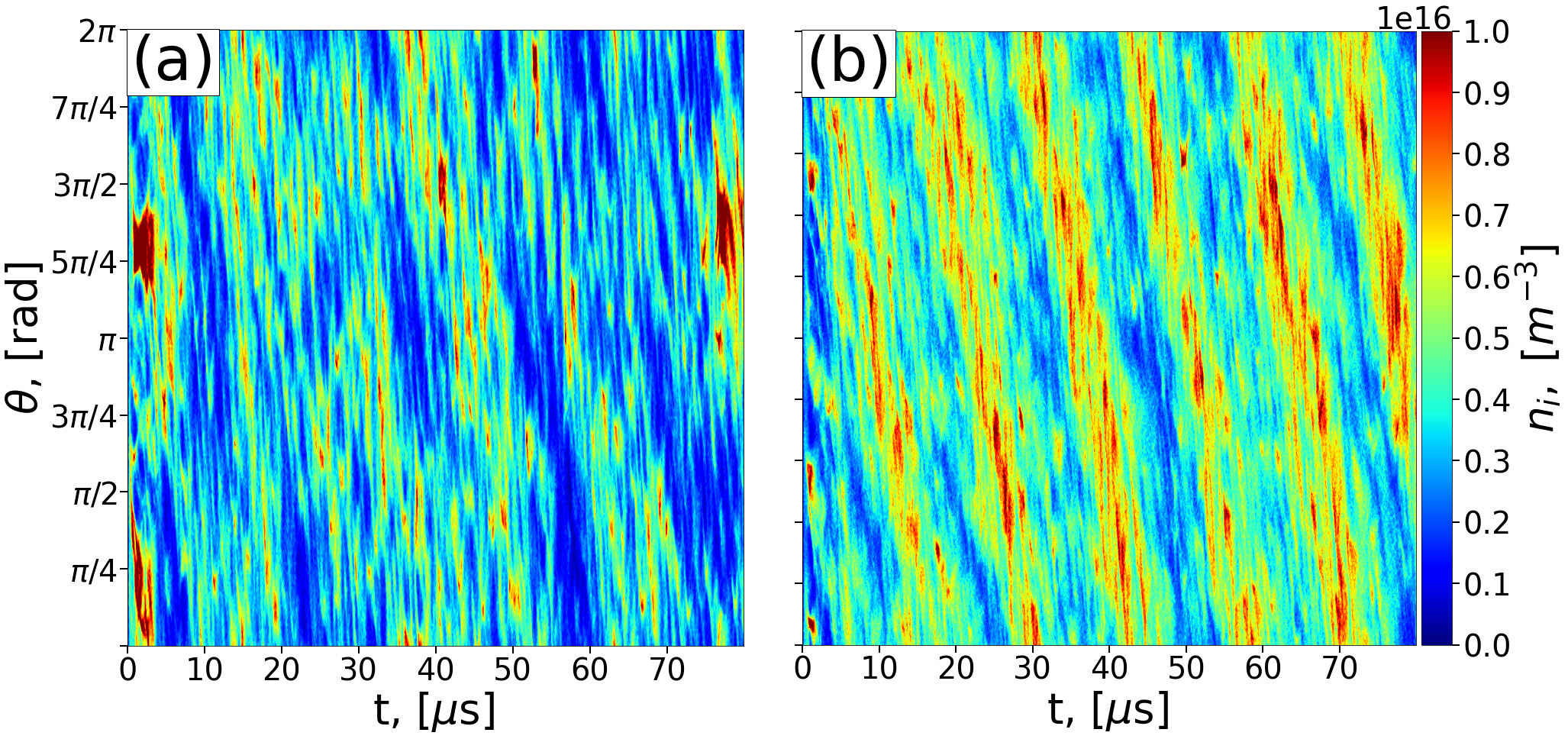}
\captionsetup{justification=raggedright,singlelinecheck=false}
\caption{Angular rotation of the ion density fronts: a) the case when only ionizing collisions are included;  b) full electron neutral collisions and ionization are included.  The spoke is dispersed and less evident when only ionization is included, consistent with the result of the probe measurements in \Cref{fig:PEC2PICProbAnalysis}.}
\label{fig:TemporalSpatialIonBothCases}
\end{figure}

The role of different types of collisions on the spoke frequency was also studied  in EDIPIC-2D simulations. In \Cref{fig:TemporalSpatialIonBothCases} we show  the temporal-spatial evolution of the ion density visualizing the spoke rotation in EDIPIC-2D data. Here a circular ring is considered at a quarter of the box size (half of a radius) with a width of 10 cells. The ring is split into 800 sectors so that the extent of one sector is roughly twice the cell size. The data is averaged over each sector area and the information from all sectors is plotted as an angular dependence.  One can clearly see the propagating density fronts corresponding to the rotating spoke structure. The front angle in the $\theta -t $ plane is used to measure the rotation frequency. There is good agreement between the spoke frequency measured from PEC2PIC simulations using local density probe data, \Cref{fig:PEC2PICProbAnalysis},  and from EDIPIC-2D simulations using temporal-spatial plots of the spoke fronts, \Cref{fig:TemporalSpatialIonBothCases}. The diffused nature of the rotation frequency peak for the case with the ionization only (in absence of all other electron-neutral collisions) is evident in both \Cref{fig:TemporalSpatialIonBothCases}a and \Cref{fig:PEC2PICProbAnalysis}-bottom.
We have studied the convergence of  the base case results by performing the additional EDIPIC-2D simulations  with the higher average ppc numbers: a 85 for the case with only ionizing collisions, and 127 for ionizing and electron-neutral non-ionizing collisions. The comparison shows that in the simulations with the  higher ppc values, the  quantitative characteristics do not change drastically: difference in the averaged parameters is at most ~15 $\%$ reaching this maximum in the center of the domain. The  rotation frequency for the $m=1$ mode stays approximately the same,  since it is determined by plasma parameters in the bulk of the plasma, roughly at the two-thirds of the radius, where the differences (in plasma parameters) are  lower than ~5$\%$. We believe that such low sensitivity is due to the global nature of the $m=1$ mode  determined by plasma parameters in the central part of the discharge where the ppc number is  significantly  larger compared to the average value.




\section{Azimuthal structures and spectra}\label{AzimuthalStrSpectra}

 Here we discuss the main characteristics of the $m=1$ spoke and higher order $m>1$ modes observed in the simulations  and provide  some interpretations regarding their nature.  The simulations reported in this section have been performed in the square box $6\times 6\;\text{cm}^2$ with the  magnetic field of $B=200\;\text{G}$, particle weight is $3.1 \times 10^5$ (resulting in the ppc=117) the  other parameters are the same as in \Cref{table:PhysicParamBaseCas}.

\subsection{The m=1 (spoke) mode}
Formation in the nonlinearly saturated state of a large scale rotating structure (spoke)  is one of the most prominent features observed in our simulations. In the context of the Penning discharge geometry similar  structures were  observed  in experiments \cite{SakawaPFB1993,RaitsesIEPC2015,RodriguezPoP2019} and
 earlier simulations \cite{PowisPoP2018,CarlssonPoP2018}.

It is instructive to discuss the ion rotation in the equilibrium in absence of azimuthal perturbations. The ions are confined by the inward radial electric field. The ion rotation velocity in the equilibrium can be estimated from the radial momentum balance in the form 
\begin{equation}
-\frac{V^2_{\theta_i}}{r} = \frac{e}{m_i}[E_r + V_{\theta_i}B] - \frac{1}{m_i n} \nabla (T_i n_i)].
\label{Eq:brot}
\end{equation} 
The roots of \Cref{Eq:brot} for $V_{\theta i}$ correspond to two branches of Brillouin rotation modes exploited in $E\cross B$ filters for mass separation \cite{GueroultPOP2019}. The  Brillouin limit corresponds to the case of large $E_r>r\omega_{ci}B/4$ (in  neglect of the pressure gradient) when the equilibrium is lost. In our case, the radial electric field is negative,  ions are well confined,  and the effect of the magnetic field on ions is small, so that one has for the frequency of the stationary ion rotation the following estimate  
\begin{equation} 
\omega_i \simeq \sqrt{\frac{-e E_r}{m_i r}},\label{Eq:ion_rot}
\end{equation}
where 
$\omega_i = V_{\theta_i} / r$.
Formally, for our parameters,  the ion pressure gradient can be important, however the effects of pressure gradient in its fluid form in \Cref{Eq:brot} is not valid in the limit of large ion orbits so that the  kinetic theory has to be used. In our case, radial ion excursions  are large and comparable to the device radius. As a result, the effects of the radial  variations of density are largely smeared out for ions.   As we will discuss in more  details below, the frequency from \Cref{Eq:ion_rot} well describes the $m=1$ spoke rotation. 

\begin{figure}[H]

\subcaptionbox{\label{fig:SpokeFourPhases_IonConc}}{\includegraphics[width=1\linewidth]{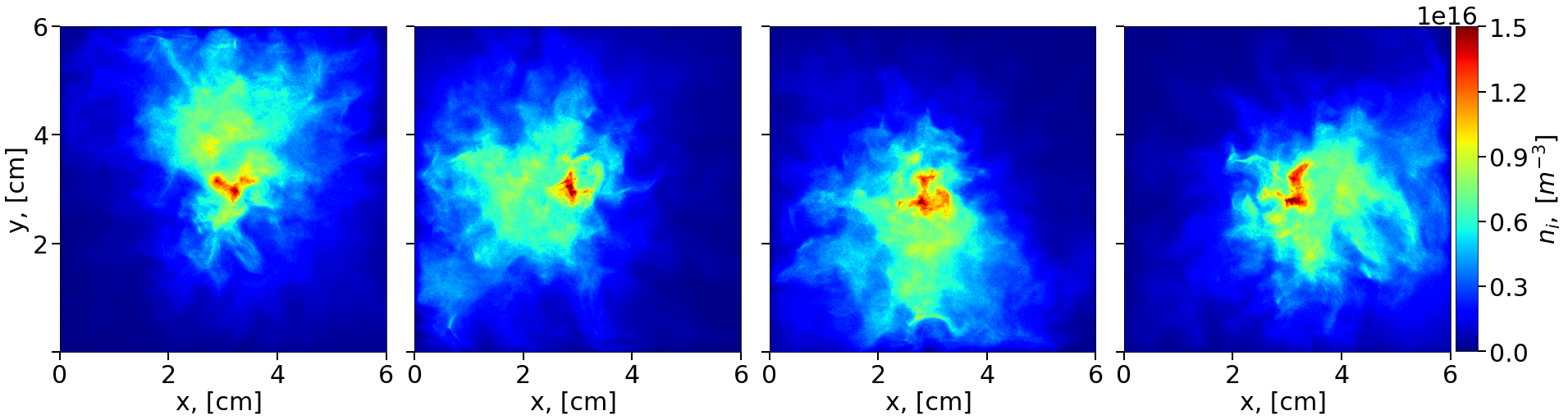}}
\subcaptionbox{\label{fig:SpokeFourPhases_Potential}}{\includegraphics[width=1\linewidth]{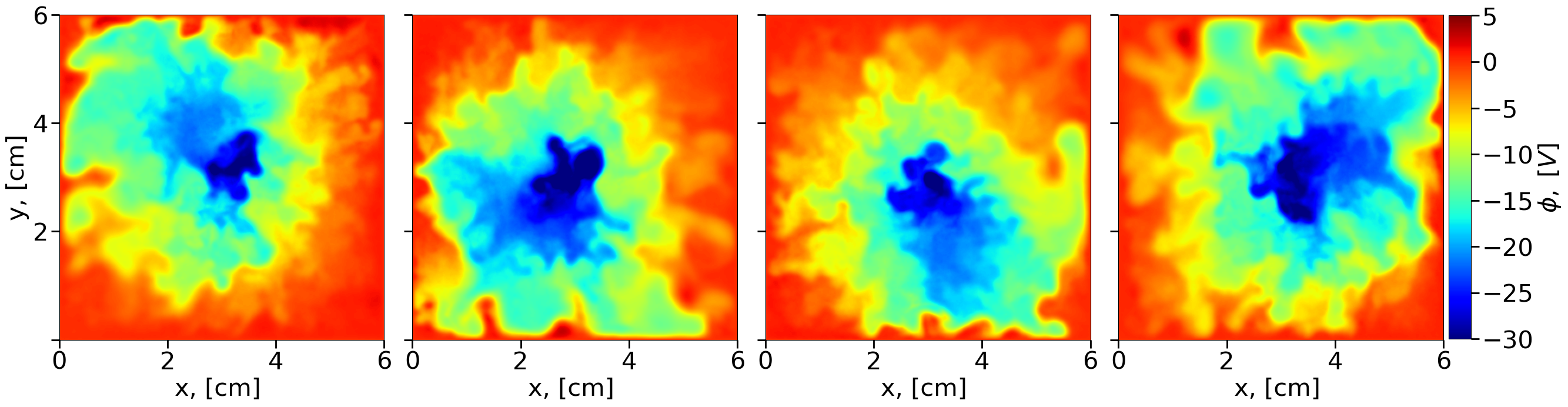}}
\subcaptionbox{\label{fig:SpokeFourPhases_Cur}}{\includegraphics[width=1\linewidth]{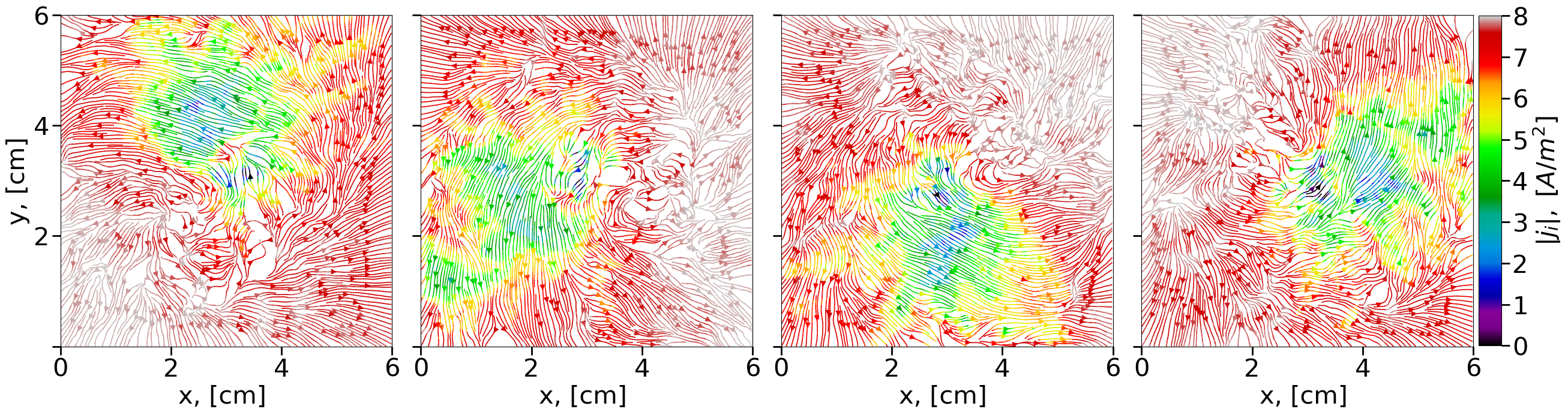}}
\centering
\captionsetup{justification=raggedright,singlelinecheck=false}
\caption{Snapshots at different times, from left to right $t=73, 79, 83, 88\;\mu s$, demonstrating spoke rotation: a) the ion density; b) the potential; c) the ion current amplitude shown by the color and streamlines plotted on a uniform grid.} 
\label{SpokeFourPhases}
\end{figure}

\Cref{SpokeFourPhases} shows the $m=1$ rotating structure in the evolution of the ion density, potential and ion current. The values of the density, potential, and current do not follow each other locally however there is good global correlation between the ion flow and potential suggesting ions are trapped by the potential. In other words, ions are globally confined by the inward radial electric  field that support the ion azimuthal rotation from \Cref{Eq:ion_rot}.  The global nature of the $m=1$ structure and ion confinement (even with the spoke present)  is  evident in the ion current flow which occurs as a result of the rotation and wobbling of the ion cloud  formed by the ionization, \Cref{fig:SpokeFourPhases_Cur}. The spoke rotation frequency remains close to the stationary ion rotation  frequency given by \Cref{Eq:ion_rot}.

\begin{figure}[htp]
\centering
\captionsetup{justification=raggedright,singlelinecheck=false}
\includegraphics[width=0.85\linewidth]{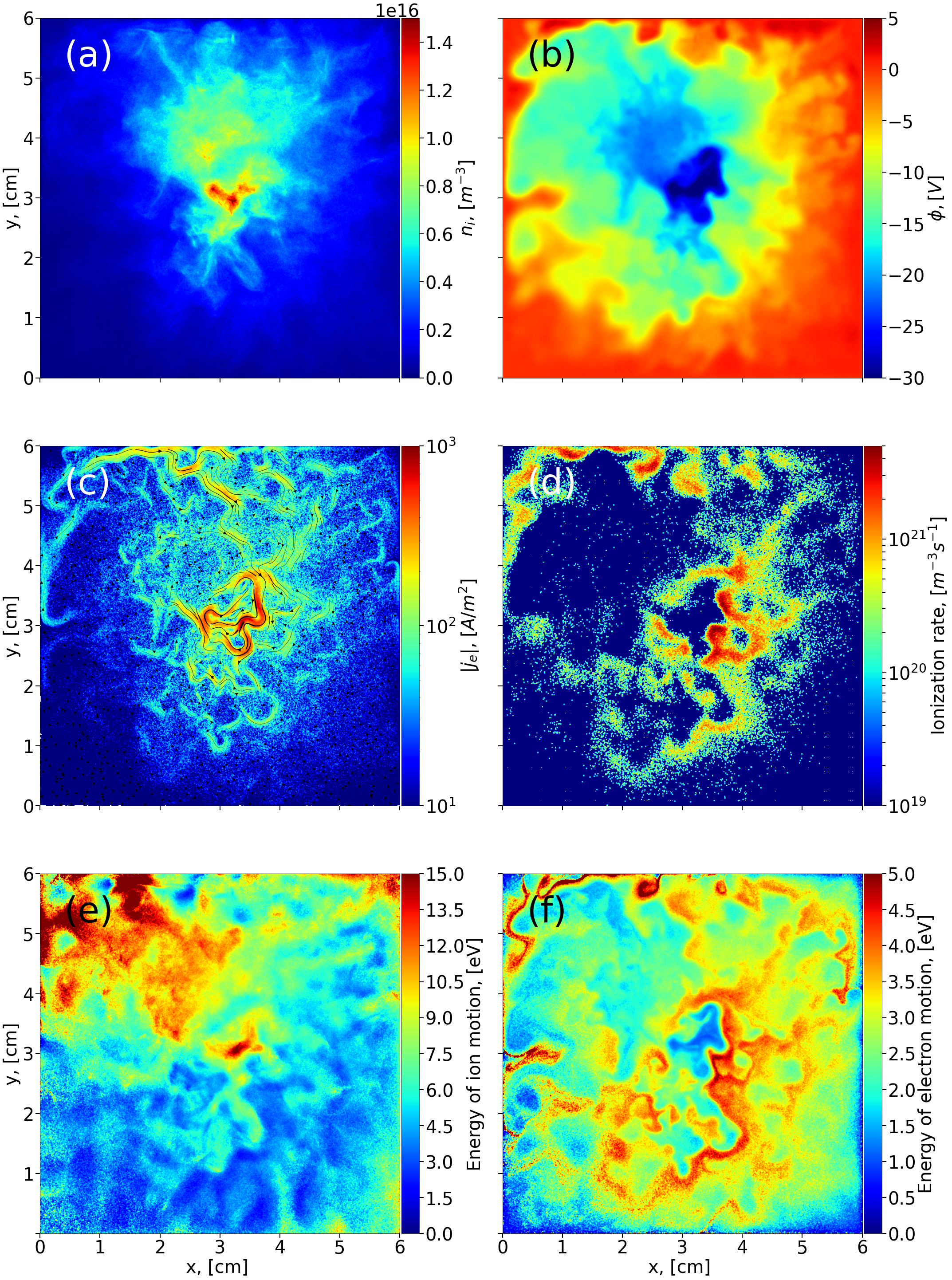}
\caption{Characteristic behavior of plasma parameters in the  spoke regime: a) ion concentration; b) potential; c) magnitude of the electron current density with streamlines (plotted on regular grid); d) ionization rate; total e) ion and f) electron energy.}
\label{fig:SpokeOneTimeSnap}
\end{figure}

Further insights on the spoke mechanism are obtained  from the  analysis of the electron current,  ion and electron energies, and ionization rate in \Cref{SpokeFourPhases}.  One can see that the electron behavior is much more local compared to that of the ions. Electrons are  locally heated in the regions of the strong electric field, at the edges of the potential structures where the gradients of the potential are large, as seen in Figures \ref{fig:SpokeOneTimeSnap}b and f. The regions of larger electron energy  are well correlated with the regions where the ionization is most pronounced, Figures \ref{fig:SpokeOneTimeSnap}b, d and f.  The electrons current concentrates inside narrow channels along the edges of the structures with large electron energy and enhanced ionization, Figure \ref{fig:SpokeOneTimeSnap}c. 

The ionization rates shown in \Cref{fig:SpokeOneTimeSnap}d were directly calculated in the code. Comparison with the the ionization rates calculated for the Maxwellian distribution via the characteristic temperatures from \Cref{fig:SpokeOneTimeSnap}c  are order of magnitude higher, thus suggesting that the tail of the electron distribution function are depleted compared to the Maxwellian.  On the other hand, the actual average elastic collision frequency  is close (within of 8\%) to the one calculated for the Maxwellian distribution.

It is interesting to note that the pattern of the ion energy distribution seems inverse to the distribution of the electron energy: the regions of the larger ion energy correlate with the regions of the lower electron energy \Cref{fig:SpokeOneTimeSnap}e and f.   If they had the same magnitude the sum of their energy would give almost uniform total energy. 
The ion energy is roughly equal to the energy of the ion stationary rotation,  $T_i \propto m_i V_{i\theta}^2/2 $, which means the ion energy is simply the kinetic energy of the oscillating ions trapped  in the global rotating $m=1$ potential structure, while the electrons are heated by collisions and local electric field fluctuations as a result of the lower hybrid type instabilities.

\subsection{Higher order modes, \texorpdfstring{$m>1$}.}
\begin{figure}[htp]
\centering
\includegraphics[width=1\linewidth]{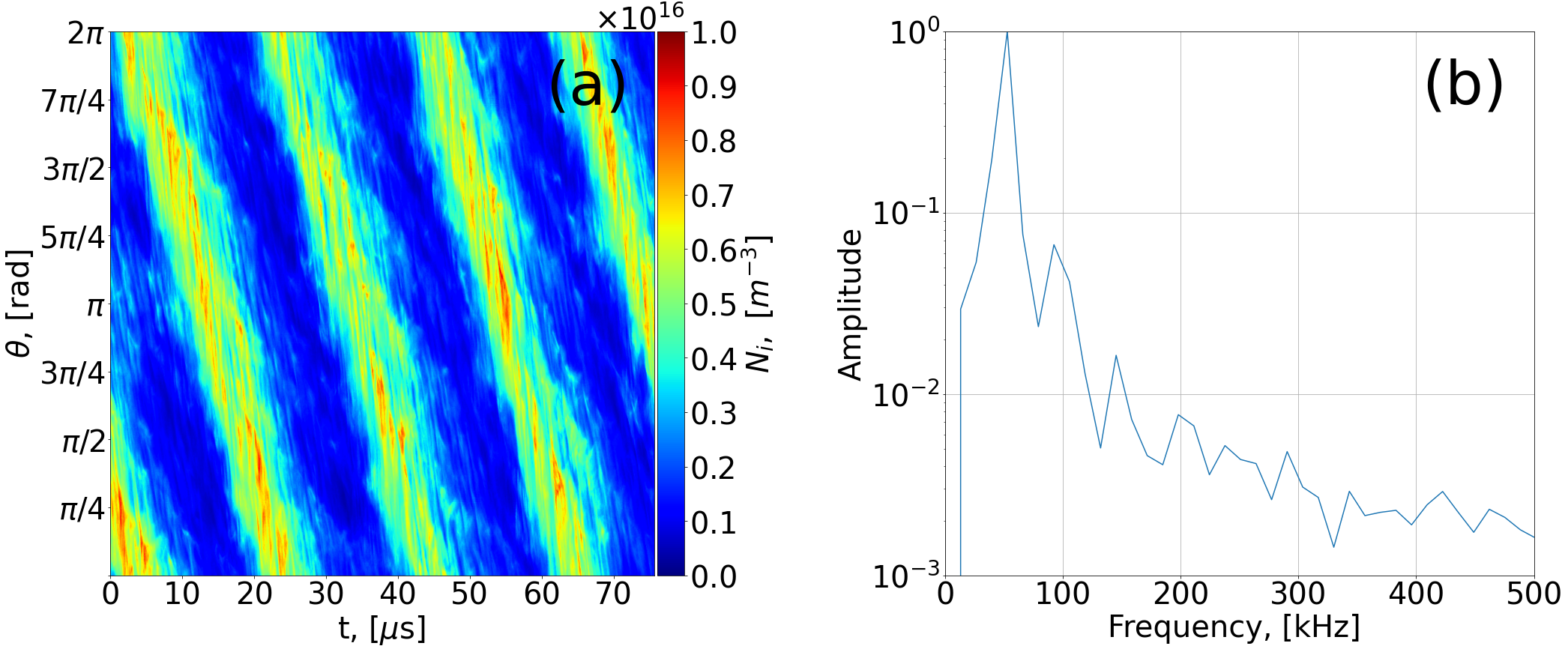}
\captionsetup{justification=raggedright,singlelinecheck=false}
\caption{Coexisting large and small scales modes. a) Ion density evolution  in the $\theta-t$ plane. The angle of the wave fronts corresponds  to the angular frequency. One can see the faster, $m>1$,  modes inside the main $m=1$ spoke structure.  b) The modes power spectrum with a range of the nonlinear harmonics.}
\label{fig:200GsAngularAndSpectra}
\end{figure}

The dominant  $m=1$ spoke structure is clearly distorted by the presence of  higher $m$ modes as seen from FFT power spectra in \Cref{fig:200GsAngularAndSpectra}b. Faster harmonics can also be seen inside  the angular-time plots of the density,  \Cref{fig:200GsAngularAndSpectra}a.  

\begin{figure}[htp]
\centering
\captionsetup{justification=raggedright,singlelinecheck=false}
\includegraphics[width=1\linewidth]{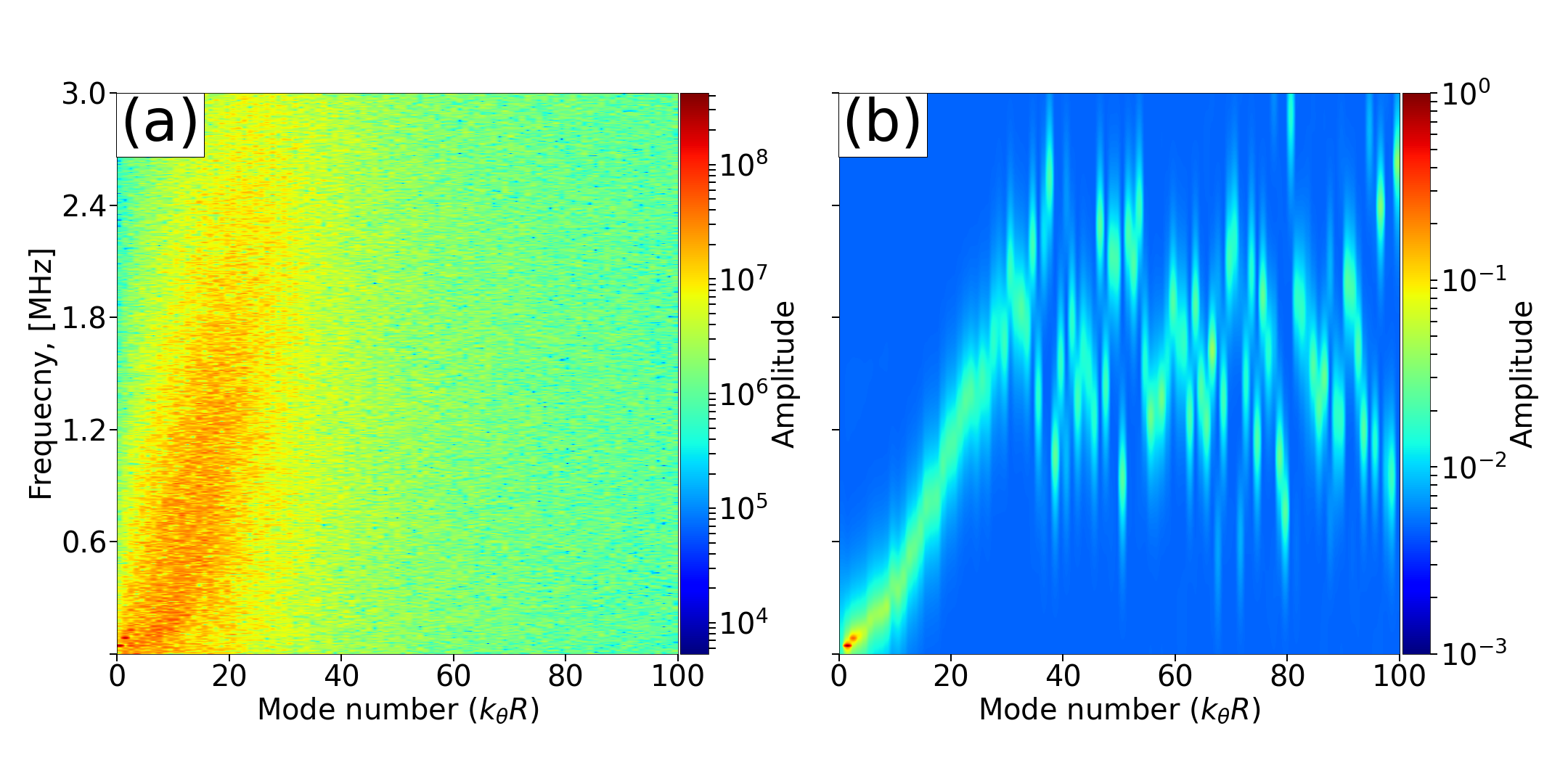}
\caption{Comparison between a) 2D FFT and b) 1D Fourier in angle variable and  MUSIC in time. The MUSIC spectra below  $m=30$ are consistent  with the 2D FFT noisy signal. The higher $m$ modes  above  $m=30$ are only seen with MUSIC. }
\label{fig:FFTversMUSIC}
\end{figure}

The two-dimensional FFT is a direct method to study  the spectra of the modes.  Such an approach, however, is difficult to apply here because it demands a much longer simulation time with many rotation periods. We use super-resolution signal processing tool -- MUSIC (multiple signal classification) which is used also in Ref.\cite{LiangPSST2021}. This method is not so strongly bound  by FFT frequency resolution limit and, most importantly, is less sensitive to noise compared to FFT. \cite{hayes2009statistical,kleiber2021modern}.  Since we have a decent resolution in angular direction, we transform each time slice in this direction with FFT, then we  apply MUSIC algorithm for each layer of $k_{\theta}$ to transform time series to the frequency space. Output of this method is an array of frequencies and the main drawback is that the relative signal strength is not reflecting the true signal strength. \Cref{fig:FFTversMUSIC} displays the comparison between two approaches: the lower modes from both methods are in reasonable agreement, while the higher modes are seen only from  MUSIC algorithm.

We interpret the observed small scale fluctuations as a result of gradient-drift instabilities  that occur due to the radial electric field, density gradient and collisions. The general linear dispersion relation for such modes in partially magnetized plasma has been proposed in Ref. \onlinecite{SmolyakovPPCF2017}. Taking into account the electron  inertia, gyro-viscosity, and collisions one has:
\begin{equation}\label{Eq:Dispersion}
 k^2_{\bot}\lambda^2_D+\frac{\omega_*+k^2_{\bot}\rho_e^{2}(\omega-\omega_0+i\nu_{e n})}{\omega-\omega_0+k^2_{\bot}\rho_e^{2}(\omega-\omega_0+i\nu_{e n})}-\frac{k^2_{\bot}c_s^{2}}{\omega^2}=0,    
\end{equation}
Here, $\omega_*$ is the diamagnetic drift frequency, $\omega_*=k_{\theta}v_*=T_ek_{\theta}/eB_0L_n$, $L_n$ is the gradient length given by $L^{-1}_n=-n^{\prime}_0/n_0$, $\omega_0=k_{\theta}v_E=-k_{\theta}E_{0r}/B_0$ is drift frequency, $k_{\theta}$ is angular wave vector in the azimuthal ($\theta$) direction, $k_r$ is the radial wave vector, $ k^2_{\bot}=k^2_{\theta}+k^2_r$, $\rho_e$ is the electron Larmor radius, $c_s$ is the ion sound speed, $E_{0r}$ is the radial electric field,  and $\nu_{e n}$ is the electron-neutral collision frequency. 

The dispersion (\Cref{Eq:Dispersion}) also includes the finite Debye length effects, given by the term $k_\perp ^2 \lambda_D^2 $, due to charge separation (non-quasineutrality).  In the long-wavelength limit,  $k_{\theta}\rho_e<<1$,  one recovers from \Cref{Eq:Dispersion} collisionless Simon-Hoh instability \cite{SakawaPRL1992} which occurs for $\mathbf{E}\cdot \nabla n_0>0$. For shorter wavelengths, the electron inertia results in the lower-hybrid mode that can be destabilized \cite{SmolyakovPPCF2017} by density gradients, $\mathbf{E}\cross \mathbf{B}$ drift, and collisions. For larger $k_{\theta}\rho_e\geq 1$ the electron response becomes Boltzmann like and one has the ion sound type mode propagating perpendicular to the magnetic field 
\begin{equation}
\omega^2=k_\perp ^2 c_s^2/(1+ k^2_{\bot}\lambda^2_D).
\label{Eq:sis}
\end{equation} It is important to note that in the limit of large collision frequency, $\nu_{en} k_{\theta}\rho_e \gg (\omega_0, \omega_*)$, the electron response (the second term in (\Cref{Eq:Dispersion})), is also Boltzmann like
and the mode again reduces to the ion sound.    In the limit of  large   $k_\perp ^2 \lambda_D^2>1$, the electron density perturbations become small and one recovers the short wavelength ion sound mode: $\omega \rightarrow \omega_{pi}$, for $k_\perp ^2 \lambda_D^2>1$, where $\omega_{pi}=\sqrt {4\pi e^2 n_0/m_i}$. 

In collisionless limit, neglecting electron inertia,  one recovers from \Cref{Eq:Dispersion} collisionless  Simon-Hoh instability \cite{SakawaPFB1993,SmolyakovPPCF2017} 
\begin{equation}
  \omega = \frac{k_{\perp}^2c^2_s}{2\omega_*} + \sqrt{ \frac{k_{\perp} ^4c^4_s}{4\omega^2_*} 
- \frac{k_\perp ^2c^2_s}{\omega_*}\omega_0}. 
\label{Eq:sh}
\end{equation}
For our parameters, the first term is small and the instability is almost aperiodic with the growth rate
\begin{equation} \label{Eq:growth}
  \gamma =  \sqrt{\frac{k_\perp ^2c^2_s}{\omega_*}\omega_0} = k\sqrt{\frac{eE_rL_n}{m_i}}.   
\end{equation}
 
It was pointed out  previously that this expression roughly correspond to the spoke rotation  frequency observed in simulations in Ref.\onlinecite{PowisPoP2018}. Note that for $k\simeq 1/r$ and $L_n\simeq r$, the \Cref{Eq:growth} gives the same estimate as \Cref{Eq:ion_rot}. \Cref{Eq:Dispersion,Eq:sh} do not include ion equilibrium rotation. When  the ion flow $V_{i \theta}$ in the equilibrium is included, the frequency in the ion response part has to be modified by the Doppler shift $\omega \rightarrow \omega -k_\theta V_{i \theta}$. Such modifications of Simon-Hoh instability were considered in Refs. \onlinecite{SakawaPFB1993,GueroultPOP2017}. 
 
\begin{figure}[htp]
\centering
\captionsetup{justification=raggedright,singlelinecheck=false}
\captionsetup[subfigure]{labelformat=empty}
\subcaptionbox{\label{fig:SpectrumB_200}}{\includegraphics[width=0.49\linewidth]{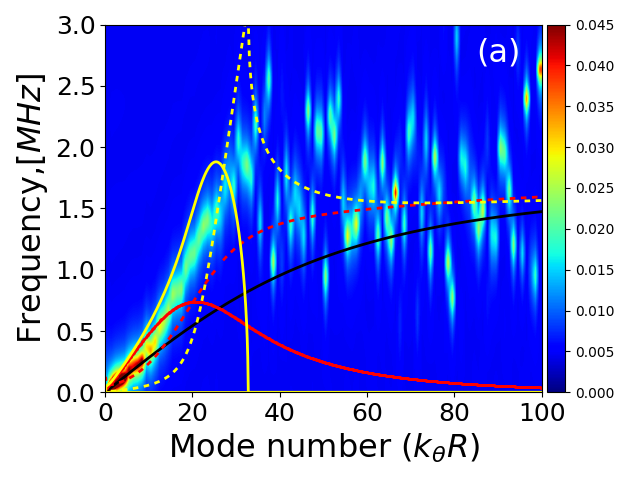}}
\subcaptionbox{\label{fig:SpectrumB_50}}{\includegraphics[width=0.49\linewidth]{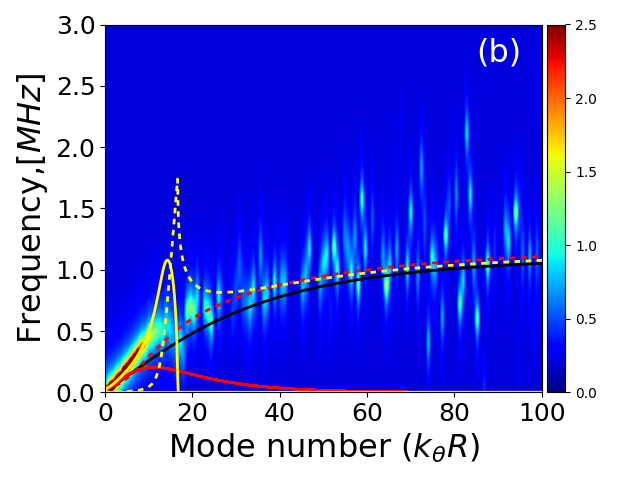}}
\caption{The 1D FFT+MUSIC transform of the space-time data of $E_{\theta}$ from simulations. Theoretical values of the real and imaginary frequency from \Cref{Eq:Dispersion} are shown as solid and dashed lines, respectively, for collisionless $\nu_{e n}=0$  (gold), and collisional  $\nu_{e n}=1.2\times \; 10^{8}$ s$^{-1}$ (red)  cases for parameters in \Cref{table:CompareLocalValueForDiff_B}. The black line  shows the ion sound wave dispersion from \Cref{Eq:sis}: a) $B=200\;\text{G}$, b) $B=50\;\text{G}$.} 
\label{SpectrumB}
\end{figure}

For our conditions the electron-neutral collisions frequency is  much larger than the spoke rotation frequency. Therefore, to  compare  the results of numerical solution  with   the $\omega-k_{\theta}$ spectra of the  $t-\theta$ field obtained in simulations, we use full \Cref{Eq:Dispersion} with collisions and  take into account the modification of the ion response due to the ion rotation.    Such  comparison is shown in \Cref{SpectrumB} for our typical plasma parameters  listed in \Cref{table:CompareLocalValueForDiff_B}  for two values of the magnetic field: $B = 50\;\text{G}$, and $B = 200\;\text{G}$.  
With collisions, \Cref{Eq:Dispersion} predicts the instability with a real part of the frequency similar to the ion-sound mode.   In the formal limit $\nu_{en}\rightarrow \infty $, one recovers from \Cref{Eq:Dispersion} the ion-sound mode, as shown in \Cref{Eq:sis}. This is also shown by black line in \Cref{SpectrumB}.  
One can see that  the real part of the frequency in collisional case start to resemble the ion sound mode (shown by black line) and the  effect is stronger for lower magnetic field, as in  \Cref{fig:SpectrumB_50}.  Nevertheless, these modes have to be classified as the gradient-drift modes because the instability is caused  by density gradient and collisions. The  dissipation results in the positive feedback phase shift between the density and potential perturbations leading to the mode growth \cite{SmolyakovPPCF2017}. The growth rate in collisional case is  shown \Cref{fig:SpectrumB_200} by solid red line. 

\begin{figure}[htp]
\centering
\captionsetup{justification=raggedright,singlelinecheck=false}
\captionsetup[subfigure]{labelformat=empty}
\subcaptionbox{\label{fig:m1wo_B200}}{\includegraphics[width=0.49\linewidth]{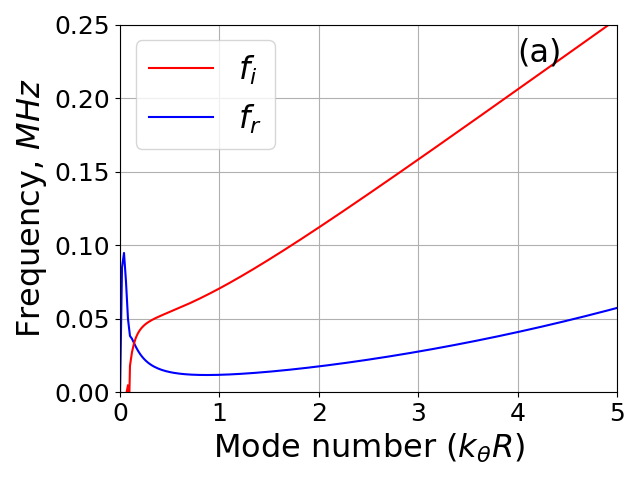}}
\subcaptionbox{\label{fig:m1wo_B50}}{\includegraphics[width=0.49\linewidth]{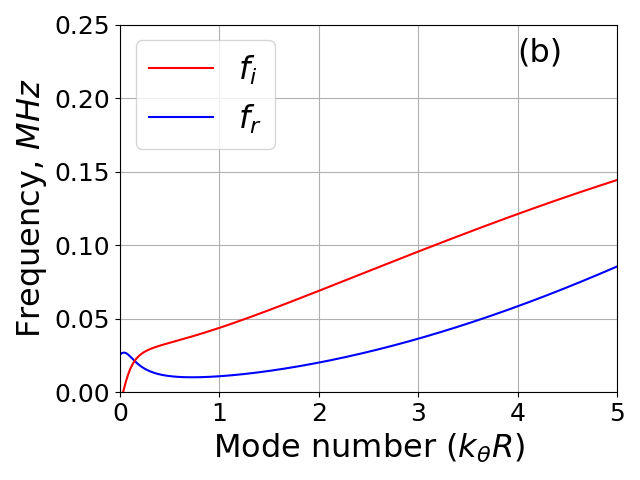}}
\caption{The low $m$ modes  linear growth rate (red) and frequency (blue)  with collisions and without the equilibrium ion rotation: a) $B=200\;G$; b)  $B=50\;G$.}
\label{m1wo}
\end{figure}

\begin{figure}[htp]
\centering
\captionsetup{justification=raggedright,singlelinecheck=false}
\captionsetup[subfigure]{labelformat=empty}
\subcaptionbox{\label{fig:m1with_B200}}{\includegraphics[width=0.49\linewidth]{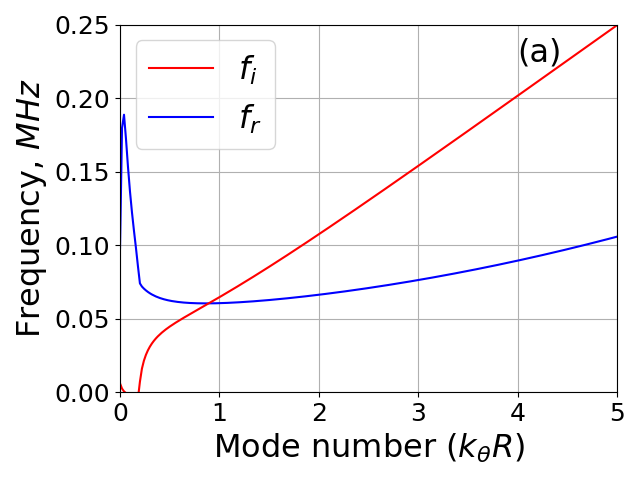}}
\subcaptionbox{\label{fig:m1with_B50}}{\includegraphics[width=0.49\linewidth]{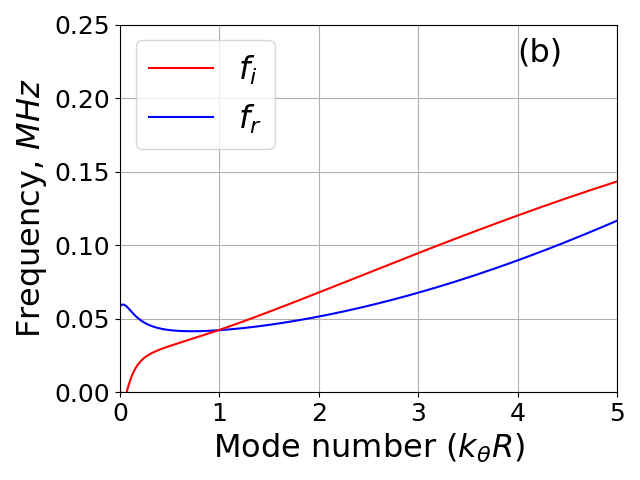}}
\caption{The low $m$ modes  linear growth rate (red) and frequency (blue)  with collisions and the equilibrium ion rotation in \Cref{Eq:ion_rot}: a) $B=200\;G$; b)  $B=50\;G$.} 
\label{m1with}
\end{figure}

Only  qualitative agreement can be expected (at best) between the results of the local linear theory with nonlinear spectra of oscillations in the saturated state. Nevertheless, general trends are in agreement with predictions based  on the gradient-drift instabilities from \Cref{Eq:Dispersion}.
The lower $m$  modes are less affected by finite collisionality though it remains important, especially for the lower magnetic field as \Cref{fig:SpectrumB_50}.   Increasing collision frequency shifts the real part of the frequency to a higher value while decreasing the growth rate. This provides the justification for the result in \Cref{fig:PEC2PICProbAnalysis} that shows that including non-ionizing electron neutral interaction increases the rotation frequency of the spoke and makes the peak wider.    
 The  equilibrium ion rotation also affects the low $m$ modes as shown in \Cref{m1wo,m1with}. The ion rotation increases the real part of the frequency so that the growth rate and real part become very close to each other for $m=1$ mode, see \Cref{m1with}.

\begin{figure}[htp]
\centering
\captionsetup{justification=raggedright,singlelinecheck=false}
\captionsetup[subfigure]{labelformat=empty}
\subcaptionbox{\label{fig:InverseCas_B200}}{\includegraphics[width=0.49\linewidth]{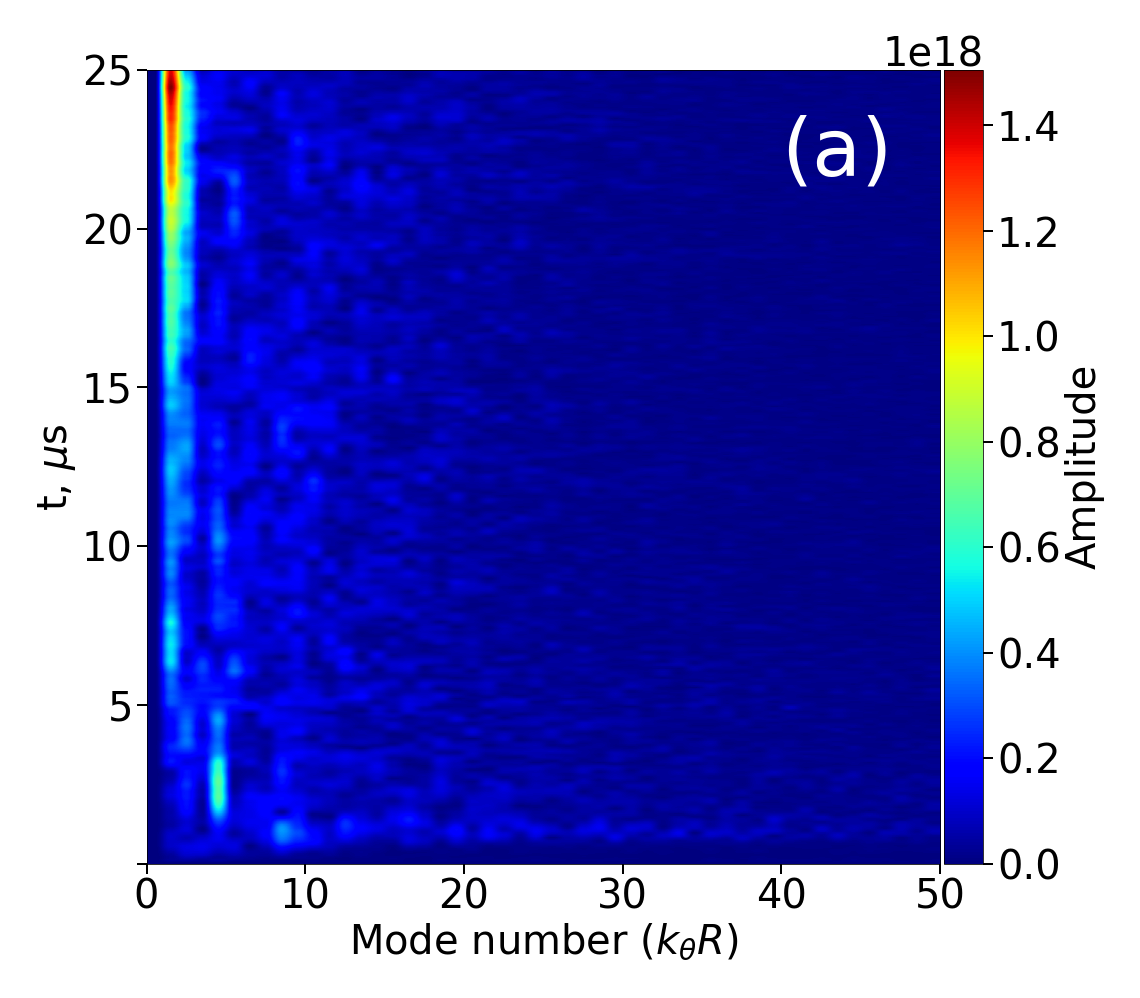}}
\subcaptionbox{\label{fig:InverseCas_B50}}{\includegraphics[width=0.49\linewidth]{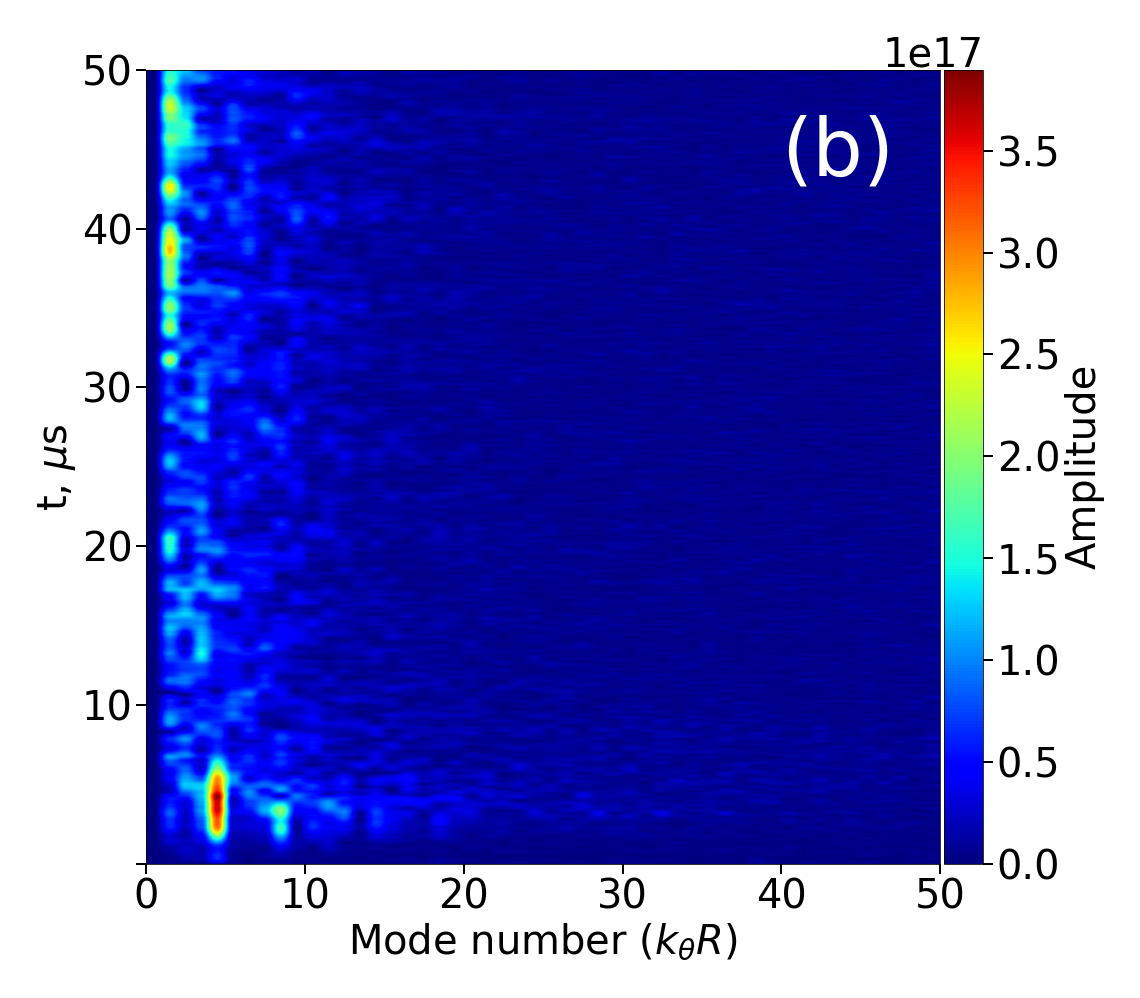}}
\caption{The azimuthal electric field $E_{\theta}$ at $r=R/2$ is transformed by 1D FFT method in the angle direction, at each time step for a) $B=200\;\text{G}$, and b) $B=50\;\text{G}$. For both cases, the  inverse energy cascade can be observed as time progresses.} 
\label{InverseCas}
\end{figure}
 The dynamics of the low $m$ modes is affected by the inverse cascade predicted for these systems earlier \cite{SmolyakovPPCF2017,KoshkarovPRL,LakhinPoP2016}.  Merging of higher $m$ modes into the the dominant $m=1$ is evident in \Cref{fig:InverseCas_B50}. The initial  m=4 mode and its m=8  harmonic correspond to the initial periodicity of the  square geometry of the simulation box. With time, these modes are reduced,  m=1 and m=2 emerge, which grow in amplitude until they reach a steady state with a dominant m=1 mode. The weak  high $m$ modes with $m>10$ also visibly reduce with time in \Cref{fig:SpectrumB_200,fig:SpectrumB_50}. Nevertheless, some activity of higher m modes remains present as shown in \Cref{fig:200GsAngularAndSpectra}a and b.


\section{Parametric study of the  m=1 spoke characteristics as function of the magnetic field, box size, and the ion species}

As it was discussed in \Cref{AzimuthalStrSpectra}, while the spoke mode originates from  the gradient-drift instability the steady state spoke rotation frequency is well approximated  by the \Cref{Eq:ion_rot} for the equilibrium ion rotation. We will use it to compare with the measured  m=1 frequency for different plasma parameters. We also show the profiles of plasma density, potential, and temperature for different cases. The radial electric field ($E_r$), the gradient length scale ($L_n^{-1} = n_p' /n_p $), plasma density ($n_p$), electron temperature ($T_e$) and electron-neutral collision frequency ($\nu_{en}$) are acquired from the ring at a half of a radius.     The electron temperature here is simply a measure of the averaged  energy and defined  as follows: $T_e = (T_x + T_y)/2$ where $T_{x,y} =m(\langle v^2_{x,y} \rangle - \langle v_{x,y}\rangle^2) $ are electron temperatures in the directions of x and y. As it was mentioned above, our convergence studies had determined that the  spoke is resilient to the noise due to the low ppc number. Comparison of the simulations with ppc=117 and ppc=11  for the $6\times6 \, cm^2$ case shows qualitatively similar behavior  while quantitatively the difference in averaged parameters is at most of the order of 5-6 $\%$ . As it was noted above, this is explained  by the fact that the actual ppc in the central part of the simulation domain is higher compared to the averaged value. In this section, the scalings for the averaged parameters and $m=1$ frequency are shown for the low resolution cases with the averaged ppc of the order of 10 (particle weight is $3\times10^6$).  
\vspace{-0.5cm}
\subsection{Effects of the the magnetic field }
\begin{figure}[htp]
\centering
\includegraphics[width=1\linewidth]{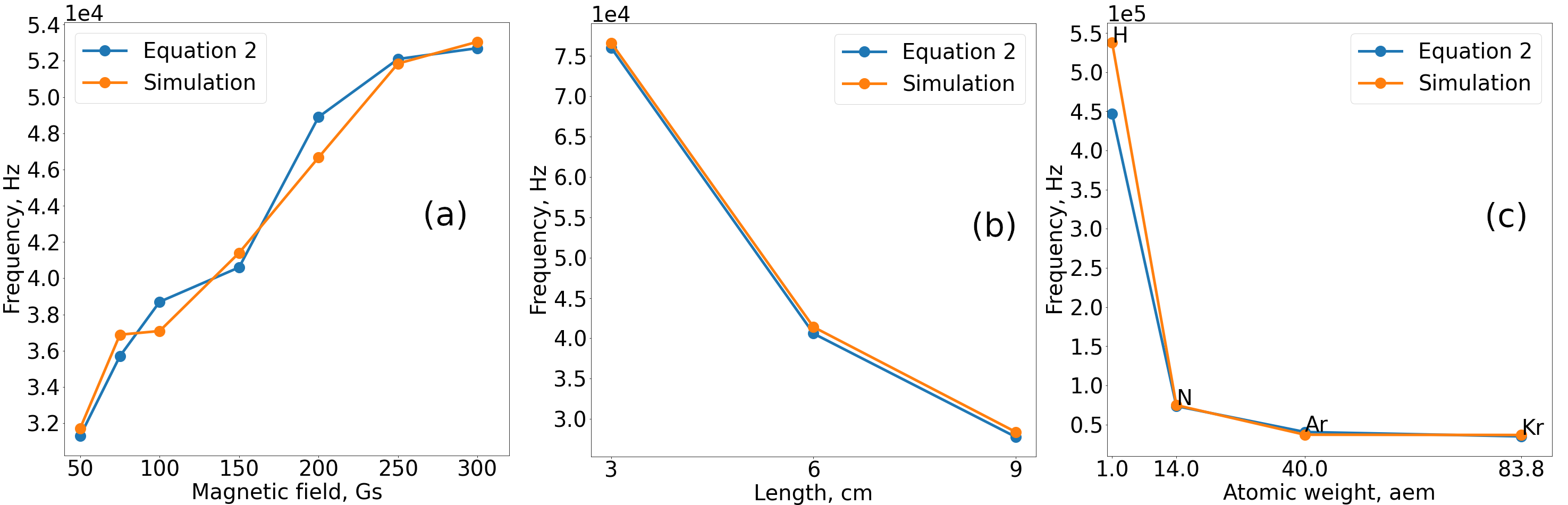}
\captionsetup{justification=raggedright,singlelinecheck=false}
\caption{The spoke rotation frequency as a function of a) magnetic field; b) size, and c) atomic element.}
\label{fig:FreqScale}
\end{figure}

\begin{figure}[htp]
\centering
\includegraphics[width=1\linewidth]{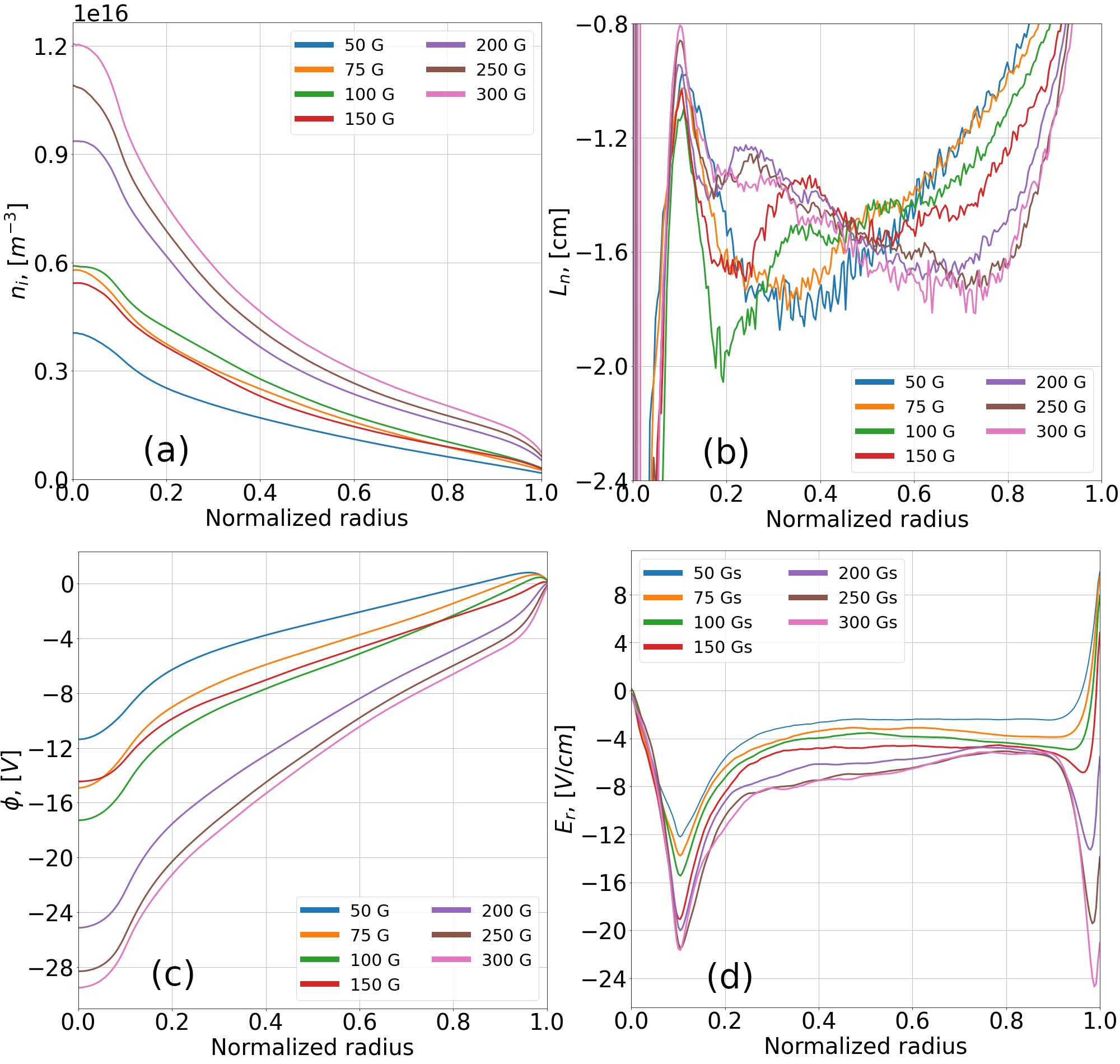}
\captionsetup{justification=raggedright,singlelinecheck=false}
\caption{Radial profiles of plasma parameters for different  the magnetic field: a) ion concentration; b) gradient length scale; c) potential; d) electric field.}
\label{fig:MagScale}
\end{figure}

Variation of the spoke frequency with the magnetic field obtained in simulations and comparison with the theoretical value from \Cref{Eq:ion_rot} is shown in \Cref{fig:FreqScale}a.  Variation of plasma parameters profiles with the magnetic field is shown in \Cref{fig:MagScale} and typical local values are summarized in \Cref{table:CompareLocalValueForDiff_B}. The magnetic field does not explicitly enter the \Cref{Eq:ion_rot}. Its effect however is manifested via the electric field dependence which is seen in \Cref{fig:MagScale}c, d, and \Cref{table:CompareLocalValueForDiff_B}. Increasing axial magnetic field improves plasma confinement therefore increasing the depth of potential well and the local radial electric field at $r=R/2$. Global density confinement of plasma density is also improving with the magnetic field,  see \Cref{fig:MagScale}a, but the local value of the gradient length scale does not change much with the magnetic field and remains around  1.5 cm which corresponds to the effective radius for this case with $3 \times 3$ cm simulation box. 
As it is shown in \Cref{fig:FreqScale}a, the spoke frequency roughly follows  the ${\sqrt B}$ scaling. This scaling was proposed  in Ref. \onlinecite{PowisPoP2018} based on the expression for the growth rate of the  Simon-Hoh instability given by \Cref{Eq:growth}. As it was explained in \Cref{AzimuthalStrSpectra}, those expression becomes similar to \Cref{Eq:ion_rot} for m=1 mode and constant $L_n\simeq r$ parameter as in the current simulations.  

\begin{table}[htp]
\caption{Local values of plasma parameters and spoke frequency $f$ for different values of the magnetic field at r = R/2}
\begin{tabular}{|l||*{7}{c|}}\hline
\rowcolor{Gray}
\backslashbox[6cm]{Quantity name}{Magnetic field, G}
&\makebox[2em]{50}&\makebox[1em]{75}&\makebox[1em]{100}&\makebox[1em]{150}
&\makebox[1em]{200}&\makebox[1em]{250}&\makebox[1em]{300}\\\hline\hline
$-E_r$, \text{V/cm}             & $2.43$ & $3.19$ & $3.75$ & $4.87$ & $5.86$ & $6.79$ & $6.87$  \\\hline
$-L_n$, cm               & $1.55$ & $1.44$ & $1.43$ & $1.55$ & $1.62$ & $1.58$ & $1.67$ \\\hline
$n_p, 10^{15}\;\text{m}^{-3}$  & $1.23$ & $1.74$ & $1.95$ & $2.37$ & $2.64$ & $2.95$ & $3.36$ \\\hline
$T_e$ eV                 & $2.68$ & $3.14$ & $3.09$ & $3.02$ & $2.99$ & $2.98$ & $3.37$\\\hline
$\nu_{en}$, 10$^8$\; s$^{-1}$& $1.01$ & $1.32$ & $1.30$ & $1.27$ & $1.26$ & $1.25$ & $1.70$ \\\hline
$f$, kHz                 & $31.7$ & $36.9$ & $37.1$ & $41.4$ & $46.7$ & $51.8$ & $53.1$\\\hline
$f_{cs}$, kHz ($c_s / r$)                 & $84.5$ & $91.4$ & $90.6$ & $89.7$ & $89.2$ & $89.1$ & $94.7$\\\hline\hline
\end{tabular}
\label{table:CompareLocalValueForDiff_B}
\end{table}

\subsection{The spoke frequency scaling with the size of the simulation box}
\begin{figure}[htp]
\centering
\includegraphics[width=1\linewidth]{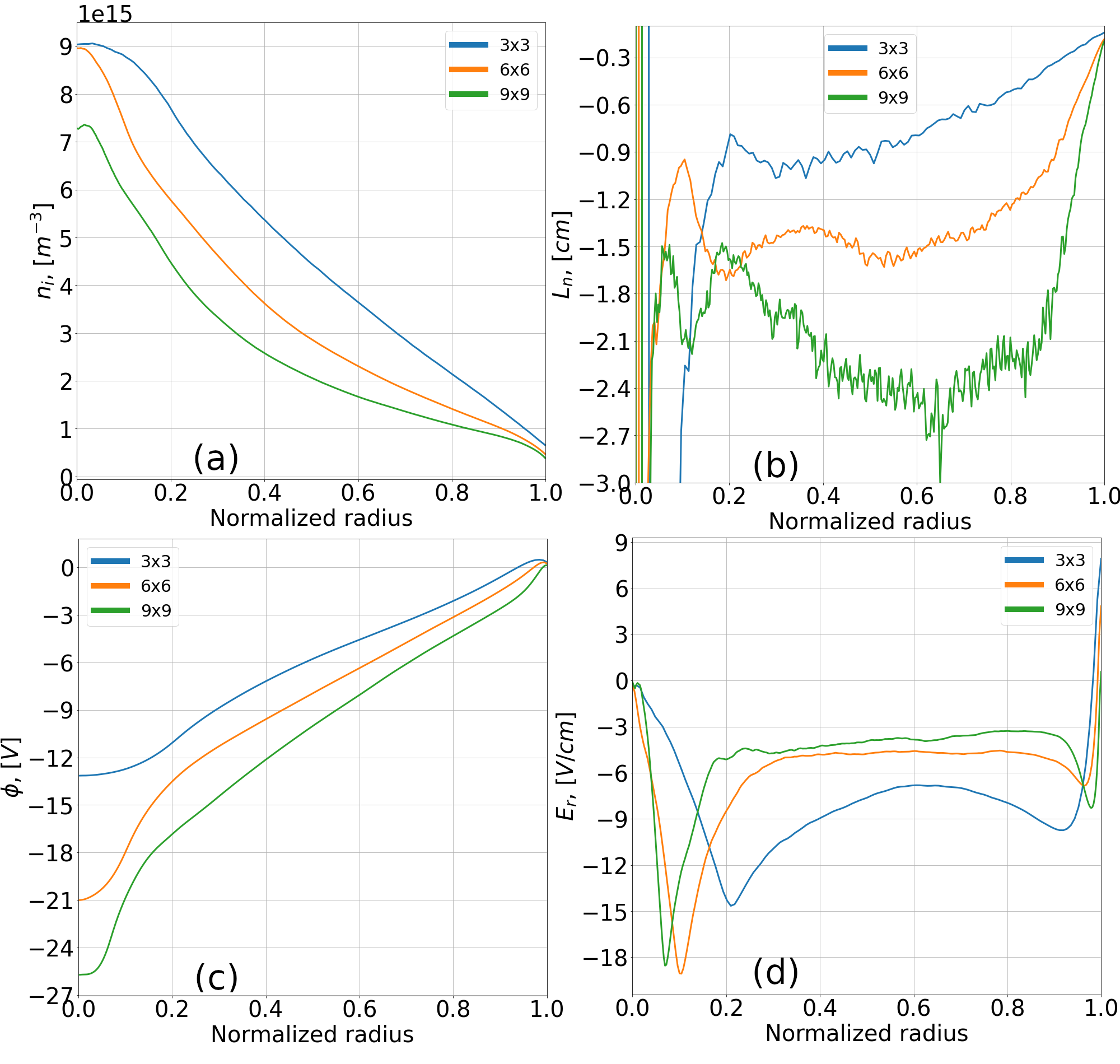}
\captionsetup{justification=raggedright,singlelinecheck=false}
\caption{Radial profiles of plasma parameters for different size of the simulation  box: a) ion concentration; b) gradient scale length; c) potential; d) electric field.}
\label{fig:SizeScale}
\end{figure}

The box size scaling shows clear dependence of the $m=1$ mode frequency as  $~R^{-1}$, see \Cref{fig:FreqScale}b suggesting  the radial electric field in the form $E_r\simeq 1/R$. The radial dependece of plasma parameters are demonstrated in \Cref{fig:SizeScale}

\begin{table}[htp]
\caption{Local values of $E_r,L_n, n_p, T_e, \nu_{en}$}
\begin{tabular}{|l||*{10}{c|}}\hline
\rowcolor{Gray}
\backslashbox[4cm]{Quantity name}{Size, $\text{cm}^2$}
&\makebox[3em]{$3\times3$}&\makebox[3em]{$6\times6$}&\makebox[3em]{$9\times9$}\\\hline\hline
$-E_r$, V/cm              & $7.13$  & $4.87$ & $3.81$ \\\hline
$-L_n$, cm                & $0.86$  & $1.55$ & $2.40$\\\hline
$n_p, 10^{15}$ $\text{m}^{-3}$   & $4.06$  & $2.37$ & $1.86$\\\hline
$T_e$ eV                  & $3.27$  & $3.02$ & $2.7$\\\hline
$\nu_{en}, 10^8$ $\text{s}^{-1}$ & $1.41$  & $1.27$  & $1.12$ \\\hline
$f$, kHz & $76.6$  & $41.4$  & $28.4$
\\\hline
$f_{cs}$, kHz & $187$  & $89.6$  & $62.2$
\\\hline\hline
\end{tabular}
\label{table:SizeScale}
\end{table}
\subsection{Effects of the ion species}
The radial profiles of plasma parameters for different elements are shown in \Cref{fig:SizeScale} and summarized in \Cref{table:AtomScale} for other parameters are from Table I.  The atomic element scaling in \Cref{fig:FreqScale}c is in good agreement with  $\sim M^{-1}$ dependence, thus  suggesting that the electric field varies with the atomic element roughly as  $\sim 1/{\sqrt M}$.  In simulations, the exact ionization energies for different elements were used, however the importance  of this factor is difficult to evaluate since the ionization energies for these elements are rather close. As it is  shown in \Cref{fig:SpokeOneTimeSnap}d,and f, there is a notable increase of the electron temperature and ionization rate at the edges of the  m=1 potential structure. This enhanced ionization however does not explain the  rotation velocity, at least not within the standard concept of the Critical Ionization Velocity phenomenon \cite{BrenningSSR1992}, in which the velocity of the ionization front is limited by the CIV value  $\sqrt{2eV_{ion}/M}$, where $V_{ion}$ is the ionization potential.       
As shown in \Cref{table:ComparisoCIVIonRotationVelocity}, the spoke rotation  velocity observed in simulations is much lower than the CIV value. We note that in all cases, the m=1 spoke rotation is much slower than the $\mathbf{E}\cross \mathbf{B}$ values.

\begin{figure}[htp]
\centering
\includegraphics[width=1\linewidth]{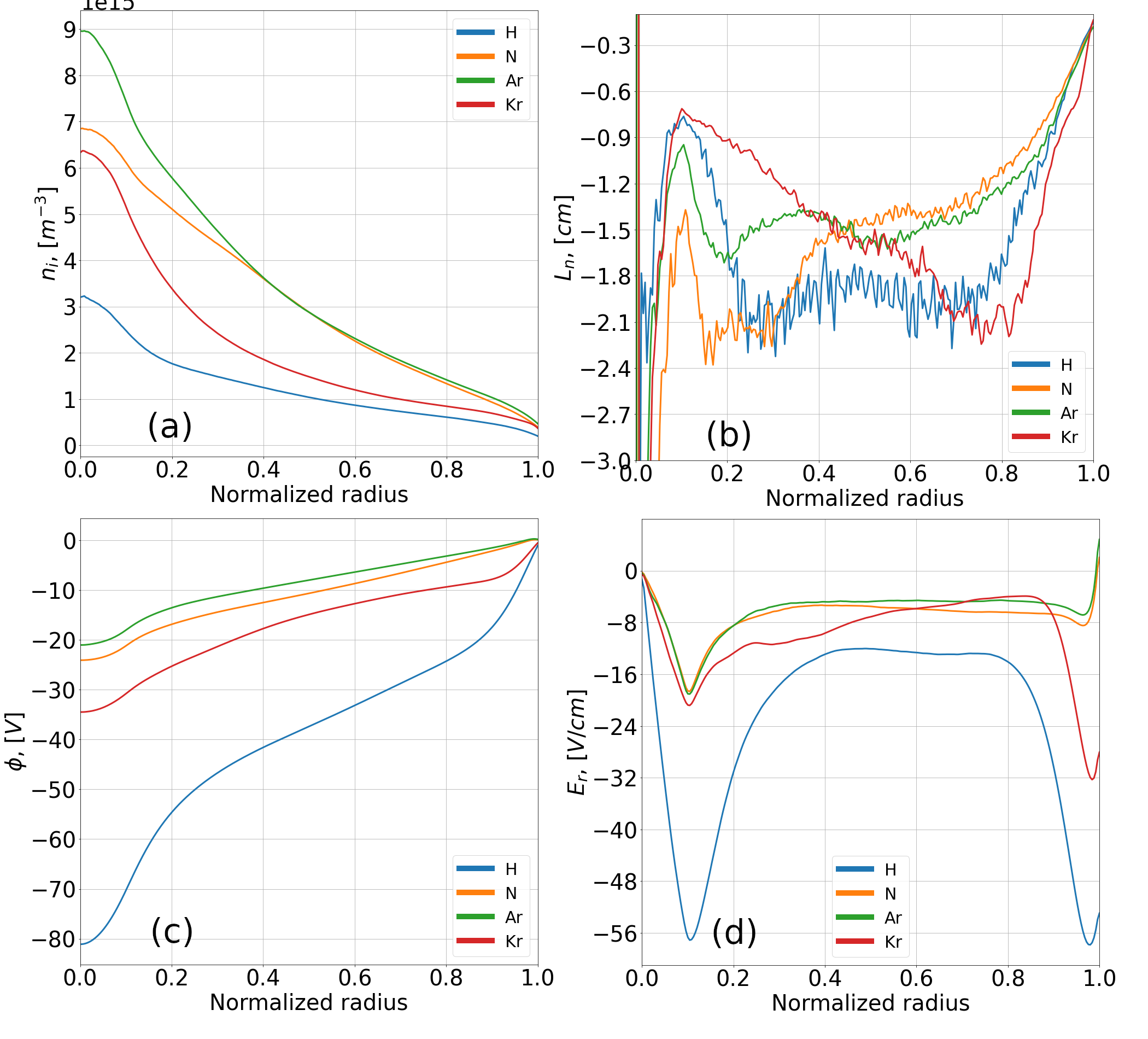}
\captionsetup{justification=raggedright,singlelinecheck=false}
\caption{Radial profiles of plasma parameters for different atomic elements:  a) ion concentration; b) gradient scale length; c) potential; d) electric field.}
\label{fig:AtomicScale}
\end{figure}

\begin{table}[htp]
\caption{Local values of $E_r,L_n, n_p, T_e, \nu_{en}$}
\begin{tabular}{|l||*{10}{c|}}\hline
\rowcolor{Gray}
\backslashbox[4cm]{Quantity name}{Element}
&\makebox[3em]{H}&\makebox[3em]{N}&\makebox[3em]{Ar}&\makebox[3em]{Kr}\\\hline\hline
$-E_r$, V/cm               & $12.35$ & $5.72$ & $4.87$ & $6.34$\\\hline
$-L_n$, cm                 & $1.91$ & $1.41$ & $1.55$ & $1.62$\\\hline
$n_p, 10^{15}$ $\text{m}^{-3}$    & $0.95$ & $2.55$ & $2.37$ & $1.33$\\\hline
$T_e$ eV                   & $3.08$ & $2.96$ & $3.02$ & $1.63$\\\hline
$\nu_{en}, 10^8$ $\text{s}^{-1}$  & $0.83$ & $9.90$ & $1.27$ & $2.93$
\\ \hline
$f$, kHz & $538$ & $74.8$ & $41.4$ & $36.7$\\\hline
$f_{cs}$, kHz & $573$ & $153$ & $89.6$ & $62.5$\\\hline\hline
\end{tabular}
\label{table:AtomScale}
\end{table}

\begin{table}[htp]
\caption{CIV and spoke  rotation velocity}
\begin{tabular}{|p{2cm}| p{2cm}| p{2cm}| p{2cm}| p{3cm}| | p{3cm}|}
\hline\hline 
\rowcolor{Gray}
\textbf{Element} & \textbf{Mass, amu} & \textbf{$\boldsymbol{V_{ion}}$, eV}
& \textbf{CIV, $10^3$ m/s} & \textbf{Spoke velocity, $10^3$ m/s} &
\textbf{$E\cross B$ velocity, $10^3$ m/s}  \\ \hline\hline
Hydrogen & $1$ & $13.6$ & $51$ & $2.57$ & $82.3$ \\\hline
Nitrogen & $14$ & $14.5$ & $14.1$ & $0.36$ & $38.1$  \\ \hline
Argon & $40$ & $15.75$ & $8.7$ & $0.2$ & $32.4$  \\ \hline
Krypton & $83.8$ & $14$ & $5.66$ & $0.18$ & $43.3$ \\  \hline \hline
\end{tabular}
\label{table:ComparisoCIVIonRotationVelocity}
\end{table}

\section{The small scale spiral arms structures regime}

In this section we investigate the transition to the regime in which small scale spiral arms azimuthal structures occur, similar to those observed in Ref. \onlinecite{LuckenPoP2019}. For our simulations  we use the box with size $6 \times 6\;\text{cm}^2$ with Argon and the magnetic field is $150\;\text{G}$, achieved ppc number is around 90, other parameters are the same as in \Cref{table:PhysicParamBaseCas}. 

In our simulations there are two mechanism of the energy input to the discharge. One mechanism  is a kinetic energy (in the axial z-direction) of injected electrons. With sufficient initial kinetic energy, such electrons may produce  ionization directly. After  scattering on neutrals, these electrons acquire a finite velocity  in the x-y plane and start to move radially as a result of further collisions and fluctuating electric field. For our base case parameters, the radial
motion of injected electrons as well as electrons and ions produced by ionization typically establishes the  radial electric field directed inward, e.g., see \Cref{fig:MagScale}c. Injected electrons diffusing radially outward create the inward electric current. In stationary state the value of this current is equal to the injection current which is fixed as an external parameter, while the radial electric field establishes self-consistently  as a result of the power and  particle balance. Co-directed electric field and current in radial direction represent the second mechanism of the energy input to the discharge. For the parameters of our  base case, exhibiting the m=1 spoke activity,  the second mechanism is dominant, as the axial energy of the injected electrons is low, see \Cref{table:ComparisoCIVIonRotationVelocity}.  For the base parameters, the axial kinetic energy of injected electron is  in fact below the ionization energy so they are unable to produce ionization directly. The electrons are heated as a result of small scale instabilities of the lower-hybrid type. Releasing cold electrons inside of the injection region with the same current produces very similar result.

\begin{figure}[htp]
\centering
\captionsetup{justification=raggedright,singlelinecheck=false}
\includegraphics[width=0.85\linewidth]{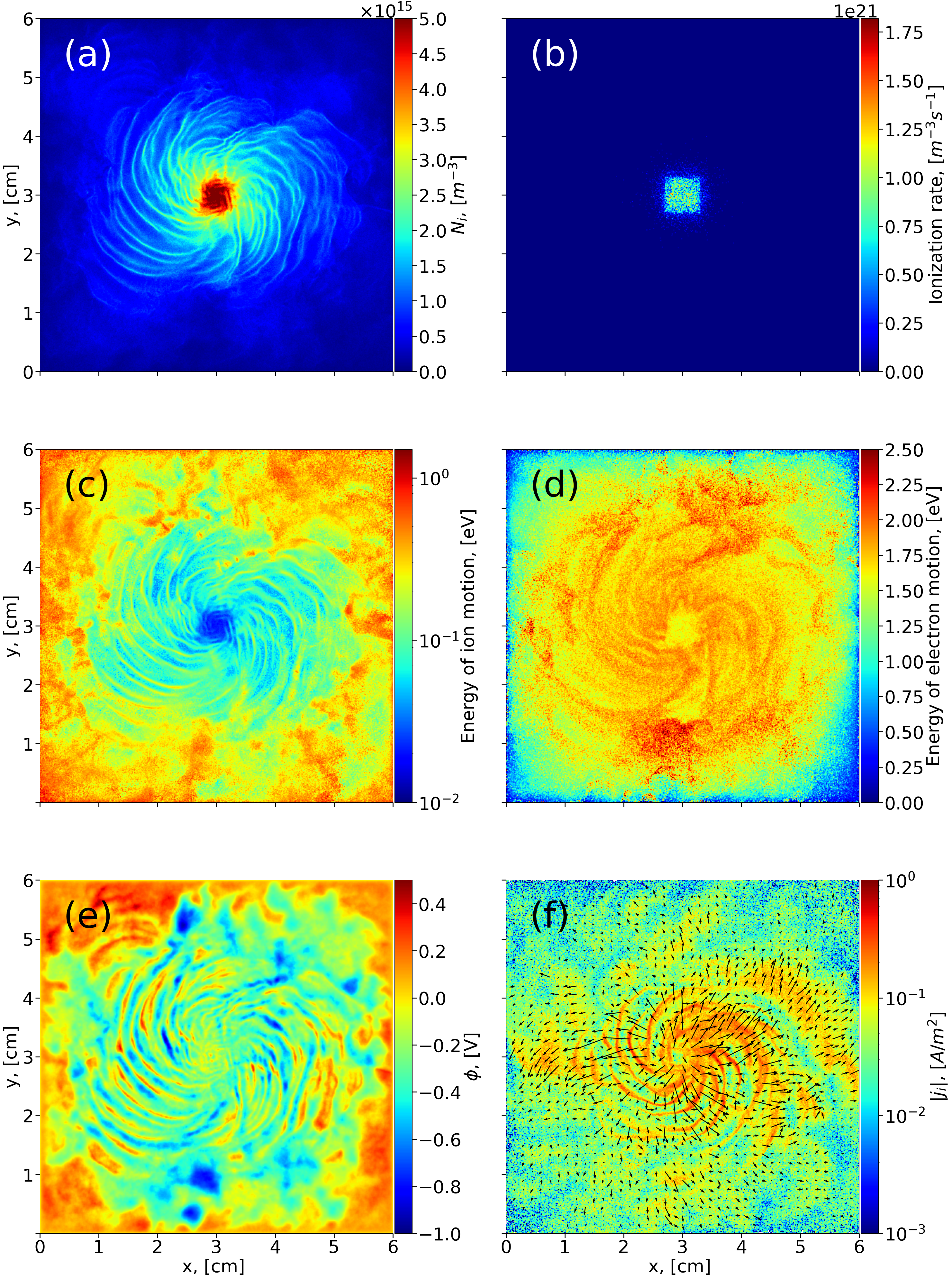}
\caption{Snapshots of plasma parameters in the small scale regime, injection energy 30 eV: a) ion concentration; b) ionization rate; c) ion energy' d) electron energy; e) potential; and f) absolute value of the ion current density with the direction vectors.}
\label{fig:SpiralSnaps}
\end{figure}

\begin{figure}[htp]
\centering
\captionsetup{justification=raggedright,singlelinecheck=false}
\includegraphics[width=1\linewidth]{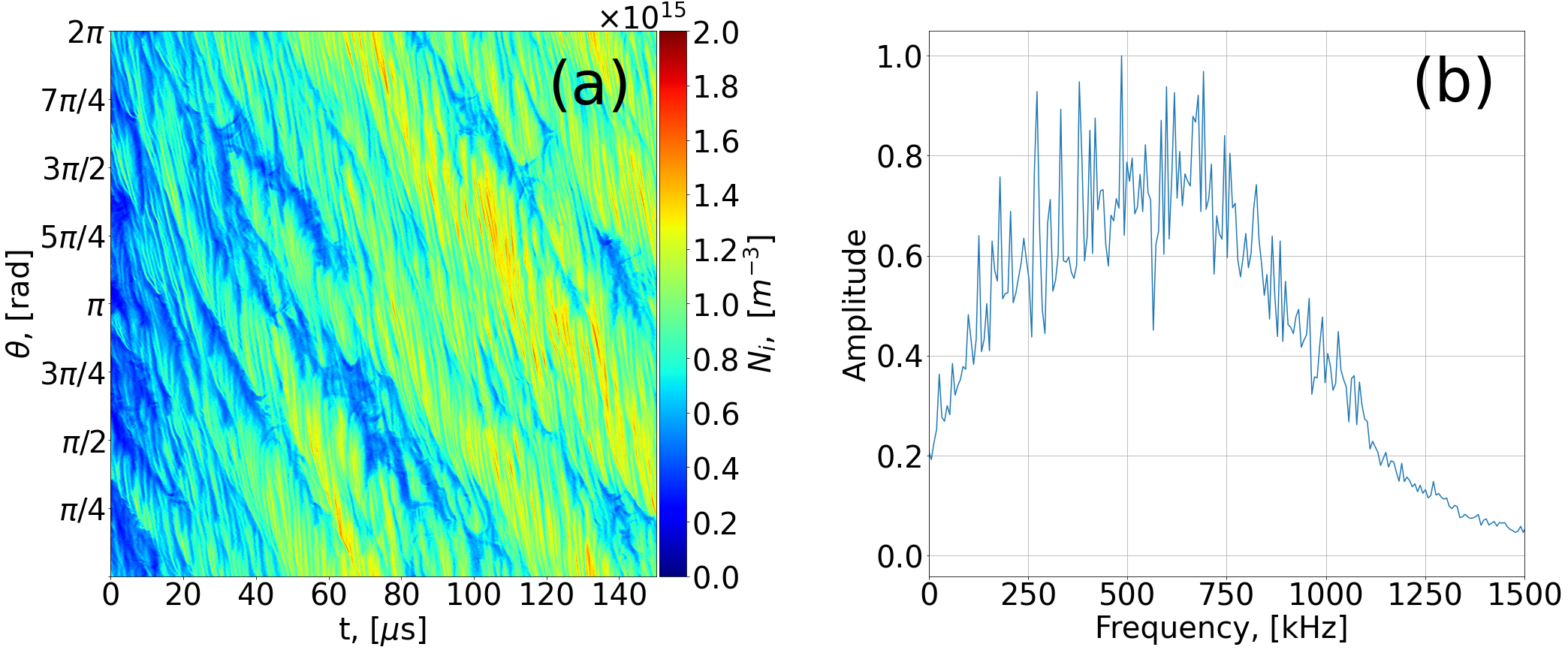}
\caption{Fluctuations spectra in the small scale regime:  a) ion density distribution in the $\theta -t$ plane; b) ion density power spectrum.}
\label{fig:SpiralSpectr}
\end{figure}

\begin{figure}[htp]
\centering
\captionsetup{justification=raggedright,singlelinecheck=false}
\captionsetup[subfigure]{labelformat=empty}
\subcaptionbox{\label{fig:SpiralComp_30ev}}{\includegraphics[width=0.49\linewidth]{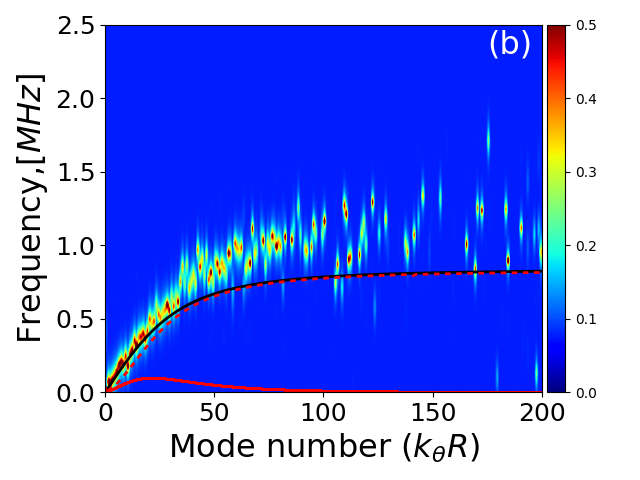}}
\subcaptionbox{\label{fig:SpiralComp_50ev}}{\includegraphics[width=0.49\linewidth]{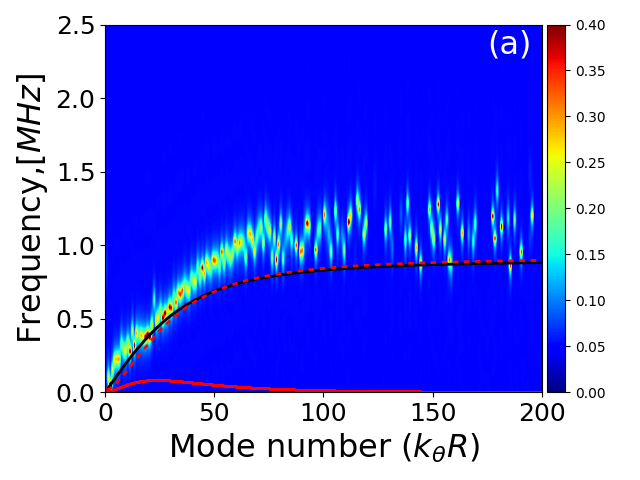}}
\caption{1D FFT+MUSIC transform of the space-time data of $E_{\theta}$.  The theoretical real part of the frequency (dashed red) and growth rates (solid red) from the dispersion relation \Cref{Eq:Dispersion} are plotted  to compare with the observed spectra. Black lines show the ion sound frequency: a) $\text{Injection energy}=30\;\text{eV}$; b) $\text{Injection energy}=50\;\text{eV}$.} 
\label{fig:SpiralComp}
\end{figure}

We observe  the disappearance  of the $m=1$ spoke activity  when the power absorption mechanism and the total power delivered to the discharge reduce. To facilitate the comparison we increase the axial beam energy to maintain roughly the same plasma density in the center. $5 \times 10^{15}\;\text{m}^{-3}$. More precisely, we perform a series of the numerical experiments  for different values of the injection kinetic energy, adjusting the injection current in each case to maintain the central plasma density near the target value. The values for the injection energy, beam current, beam power, and radial current power for these simulations are given in \Cref{table:InputPower}. As the energy delivered by the radial  current decreases,  the spoke activity is gradually reduced, the spoke becomes slower, and the system enters the regime with smaller scale, $m>1$ spiral arms structures,  as in \Cref{fig:SpiralSnaps}. The transition occurs around the injection energy of $20\;\text{eV}$.  Change of the regime is continuous, e.g., even for the case of $30\;\text{eV}$ one can see some spoke signatures  around $13\;\text{kHz}$,  \Cref{fig:SpiralSpectr}a and b. However, the power in the low frequency region corresponding to the spoke frequency is low compared to the spectral power for arms structures in the range of frequencies $400-800\;\text{kHz}$. Further increase of the energy of injected electrons  leads to even more weakened spoke and less noisy high $k$ modes  (\Cref{fig:SpiralComp}). The spoke frequency rotation is slowing down, the number of arms and their frequency is increasing: injection with 50 eV produces $7\;\text{kHz}$ and $600-900\;\text{kHz}$,  for the spoke and spiral modes,  respectively.  

\begin{table}[htp]
\caption{Input power due to the axial electron beam  and ${\bf J}\cdot  {\bf E}$}
\begin{tabular}{|l||*{10}{c|}}\hline
\rowcolor{Gray}
\backslashbox[6cm]{Quantity name}{Electron energy, eV}
&\makebox[2em]{0}&\makebox[2em]{5}&\makebox[2em]{10}&\makebox[2em]{15}
&\makebox[2em]{20}&\makebox[2em]{30}&\makebox[2em]{40}&\makebox[2em]{50}\\\hline\hline
Axial beam power, $J\, s^{-1} m^{-1}$          & $0.00$ & $0.28$ & $0.49$ & $0.48$ & $0.22$ & $0.214$ & $0.213$ & $0.212$\\\hline
$\int {\bf J}\cdot  {\bf E}dxdy $, $J\, s^{-1} m^{-1}$  & $1.92$ & $1.43$ & $0.80$ & $0.30$ & $0.02$ & $0.026$& $0.021$ & $0.03$\\\hline
Axial beam current, $A\, m^{-1}$          & $0.04$ & $0.06$ & $0.056$ & $0.03$ & $0.01$ & $0.007$ & $0.005$ & $0.004$\\\hline
\hline
\end{tabular}
\label{table:InputPower}
\end{table}

\begin{table}[htp]
\caption{Characteristic local plasma parameters for small scale regime at $r=R/2$, $B=150\;\text{G}$, for Argon.}
\begin{tabular}{|l||*{10}{c|}}\hline
\rowcolor{Gray}
\backslashbox[6cm]{Quantity name}{Electron energy, eV}
&\makebox[2em]{0}&\makebox[2em]{5}&\makebox[2em]{10}
&\makebox[2em]{15}&\makebox[2em]{20}&\makebox[2em]{30}&\makebox[2em]{40}
&\makebox[2em]{50}\\\hline\hline
$-E_r$, V/cm             & $1.41$ & $0.99$ & $0.74$ & $0.33$ & $0.31$ & $0.09$ & $0.04$ & $0.02$\\\hline
$-L_n$, cm               & $2.71$ & $2.81$ & $2.49$ & $2.64$ & $2.37$ & $2.24$ & $2.33$ & $2.39$ \\\hline
$n_p, 10^{15}$ $\text{m}^{-3}$  & $0.82$ & $0.79$ & $0.65$ & $0.51$ & $0.47$ & $0.62$ & $0.71$ & $0.76$\\\hline
$T_e$ eV                 & $2.46$ & $2.49$ & $2.46$ & $2.28$ & $1.92$ & $1.79$ & $1.74$ & $1.71$\\\hline
$\nu_{en}, 10^8$ $\text{s}^{-1}$& $1.01$ & $1.02$ & $1.01$ & $0.92$ & $0.76$ & $0.69$ & $0.67$ & $0.65$
\\\hline\hline
\end{tabular}
\end{table}

\Cref{fig:SpiralSnaps} present  the snapshots of the characteristic behavior of plasma parameters  in the small scale regime  exhibiting mutiple spiral arms structures for the injection energy of 30 eV. The plasma density, potential, ion and electron energy, and ion flow are all rather coherent and well correlated with the spiral structures. Ionization mostly occurs on the outskirts of the injection region as a ring due to the direct impact of the injected electrons.

\begin{figure}[htp]
\centering
\includegraphics[width=1\linewidth]{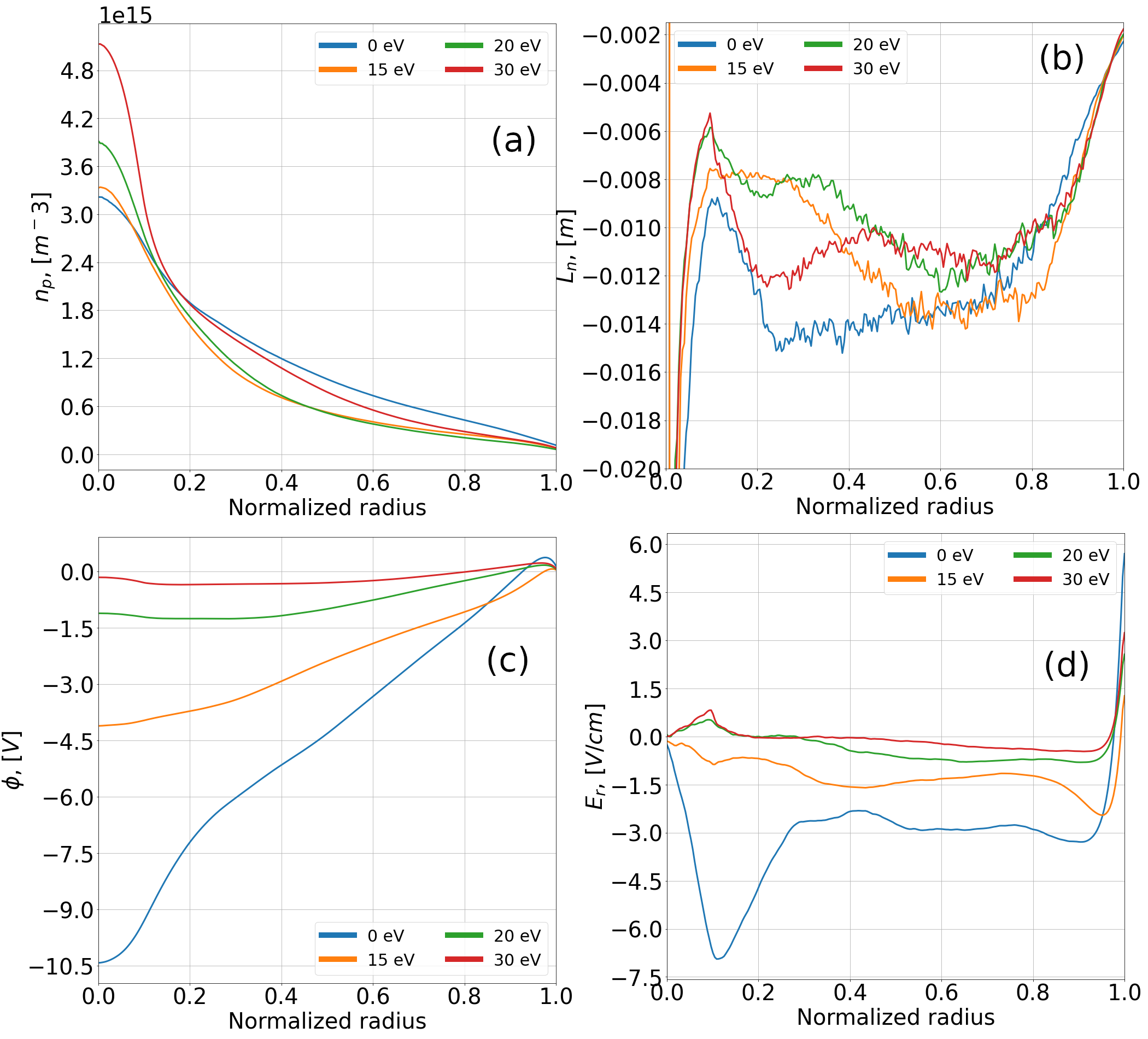}
\captionsetup{justification=raggedright,singlelinecheck=false}
\caption{Radial profiles of plasma parameters in the small scale regime for different injection energies: a) ion concentration; b) gradient scale length; c) potential; d) electric field.}
\label{fig:TransferScale}
\end{figure}

The important feature of small scale regime is a significant  reduction of the radial electric field so the  potential flattens and may even become slightly positive in the center, \Cref{fig:TransferScale}. In this case, the remaining instability mechanism is the  combination of the density gradient and collisions \cite{SmolyakovPPCF2017}. The comparison of the fluctuations spectra observed in simulations with the theoretical dispersion  from \Cref{Eq:Dispersion} is given in \Cref{fig:SpiralComp}. 
In this regime, the real part of the frequency is very close to the ion sound mode  with the growth rate defined by the density gradient and collisionality.

\begin{figure}[htp]
\centering
\captionsetup{justification=raggedright,singlelinecheck=false}
\includegraphics[width=0.5\linewidth]{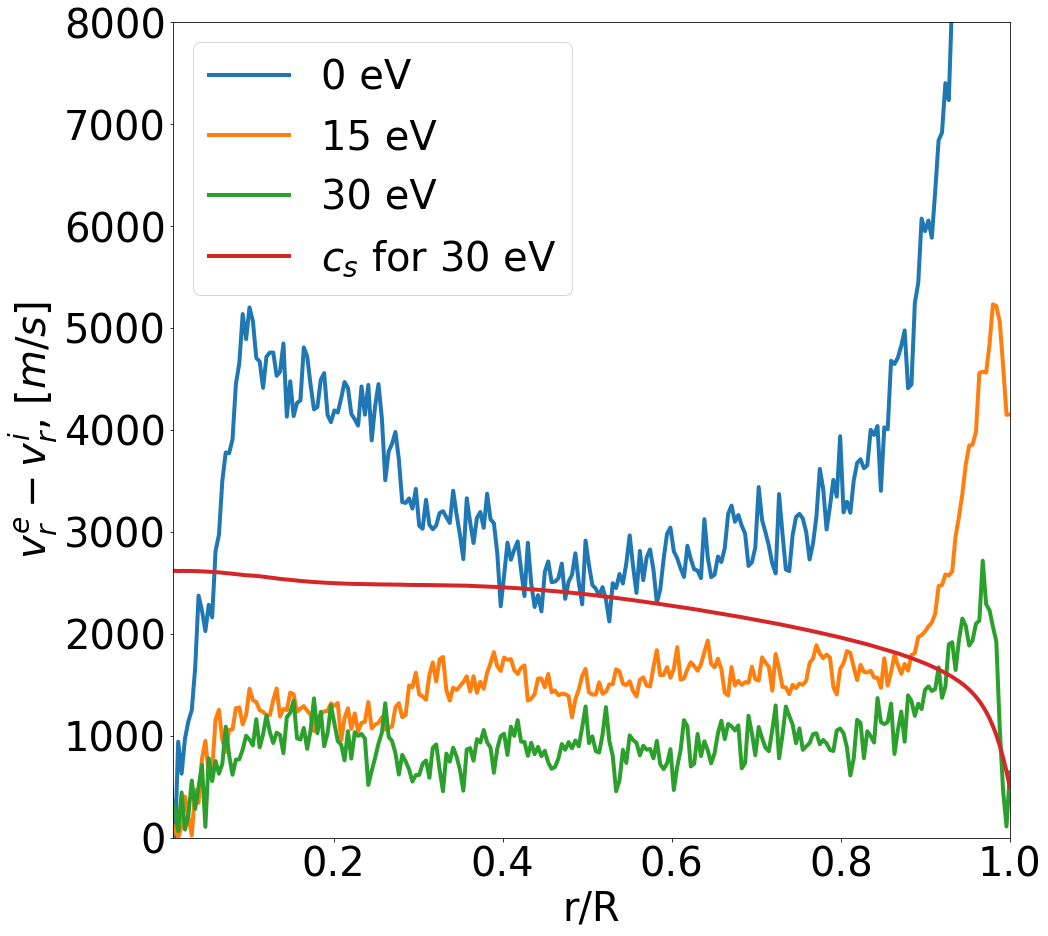}
\caption{Radial profiles of the radial current velocity  for different injection energies shown in comparison to the ion sound velocity profile. The spoke disappears for the injection energy between 15-20 eV.}
\label{fig:RadialElectronVelocity}
\end{figure}
\section{Turbulent transport and mobility.}
\begin{figure}[htp]
\centering
\subcaptionbox{\label{fig:AnomJ_B200}}{\includegraphics[width=0.47\linewidth]{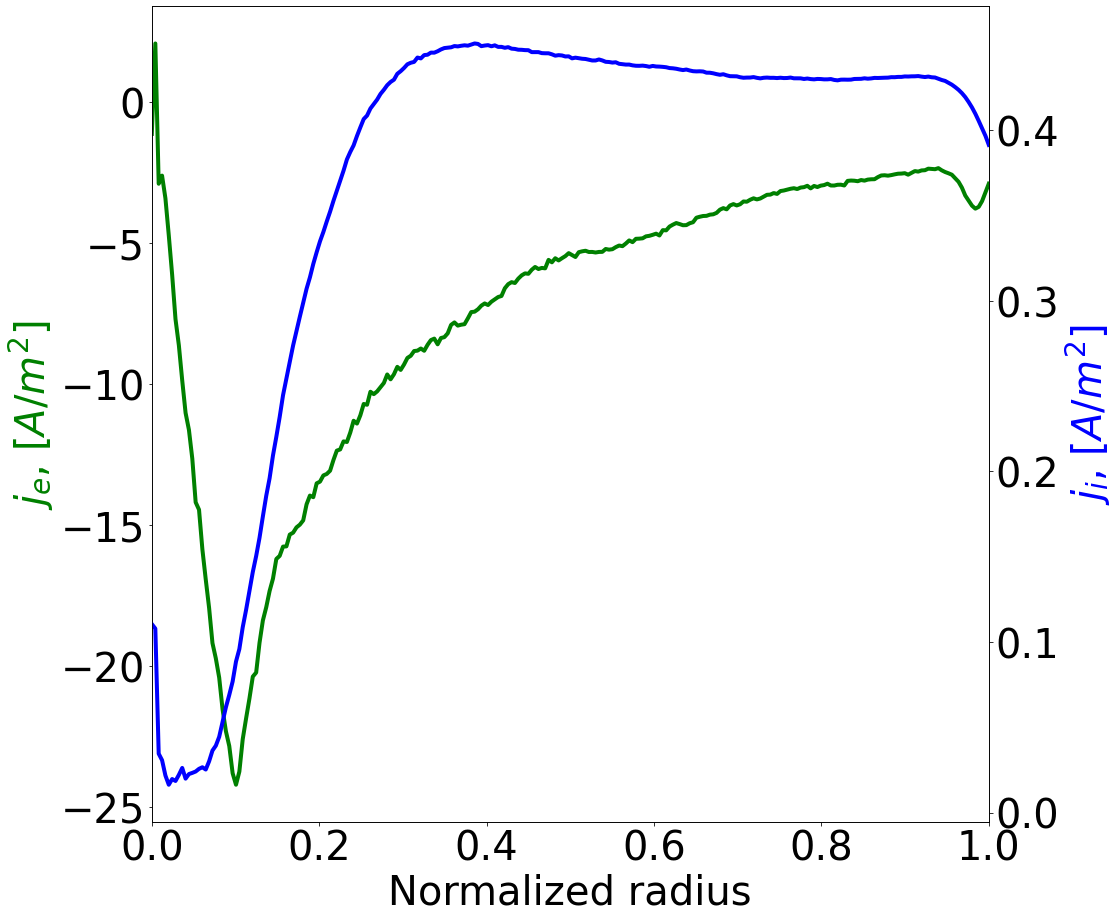}}
\subcaptionbox{\label{fig:AnomJ_B150}}{\includegraphics[width=0.49\linewidth]{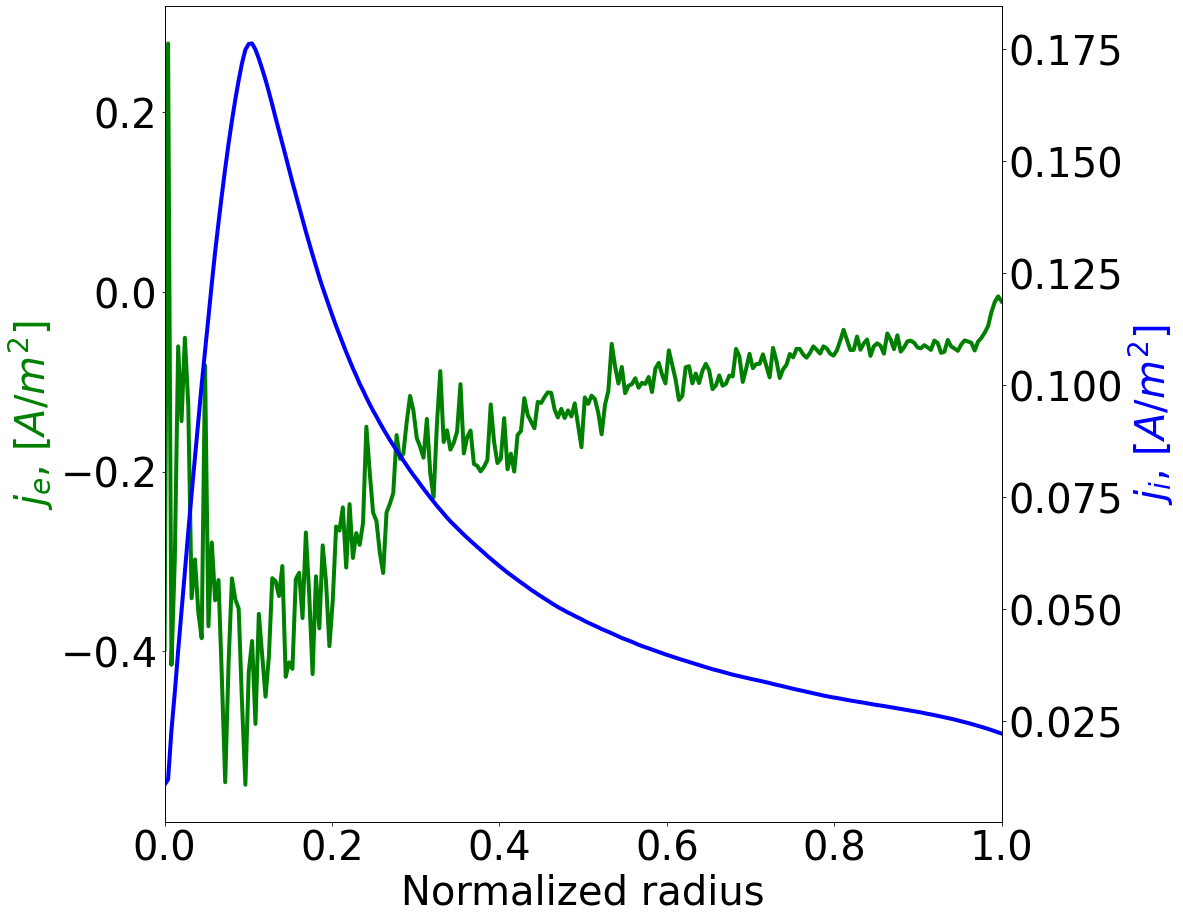}}
\caption{Electron and ion radial current: a) spoke regime, $B=200$ G; b) spiral arms regime, $B=150$ G.}
\label{fig:AnomJ}
\end{figure}

\begin{figure}[htp]
\centering
\subcaptionbox{\label{fig:egradp_B200}}{\includegraphics[width=0.5\linewidth]{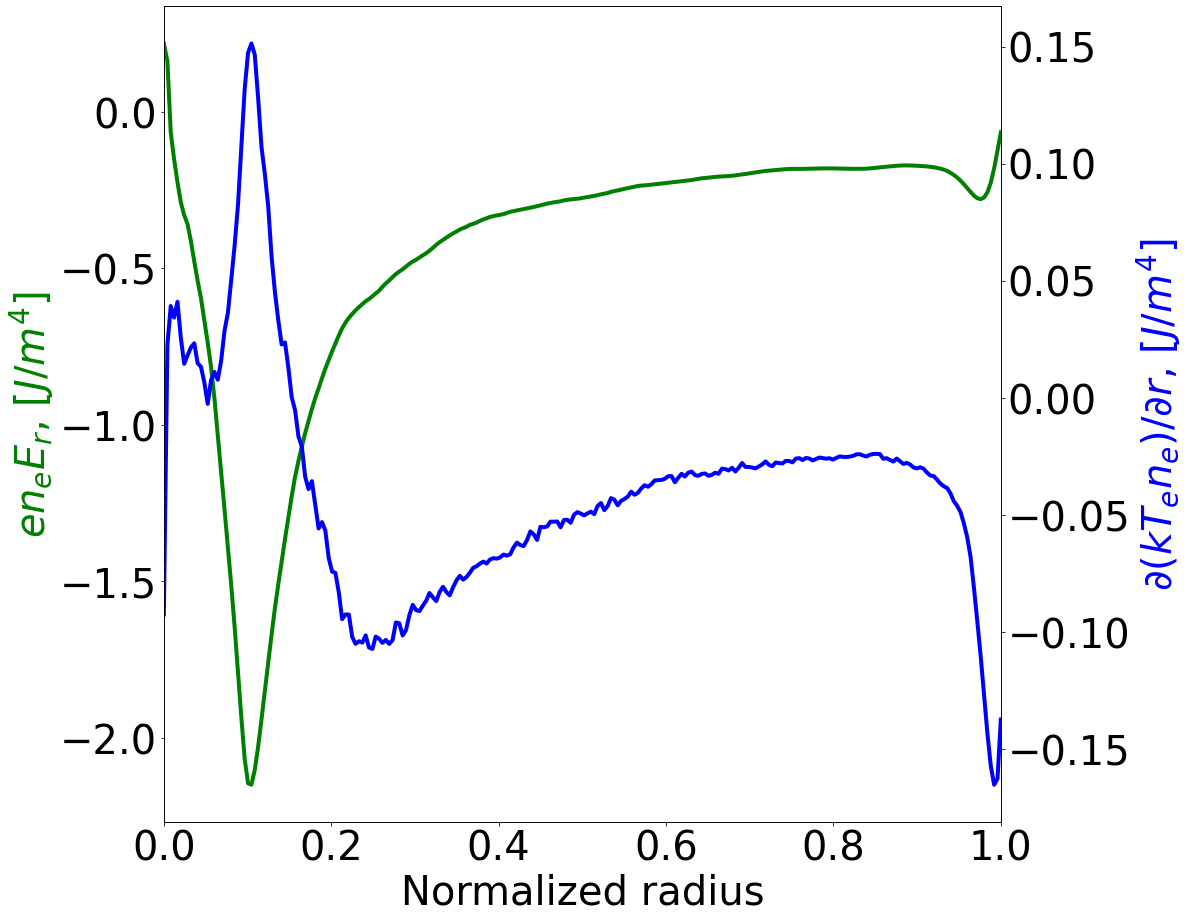}}
\subcaptionbox{\label{fig:egradp_B150}}{\includegraphics[width=0.49\linewidth]{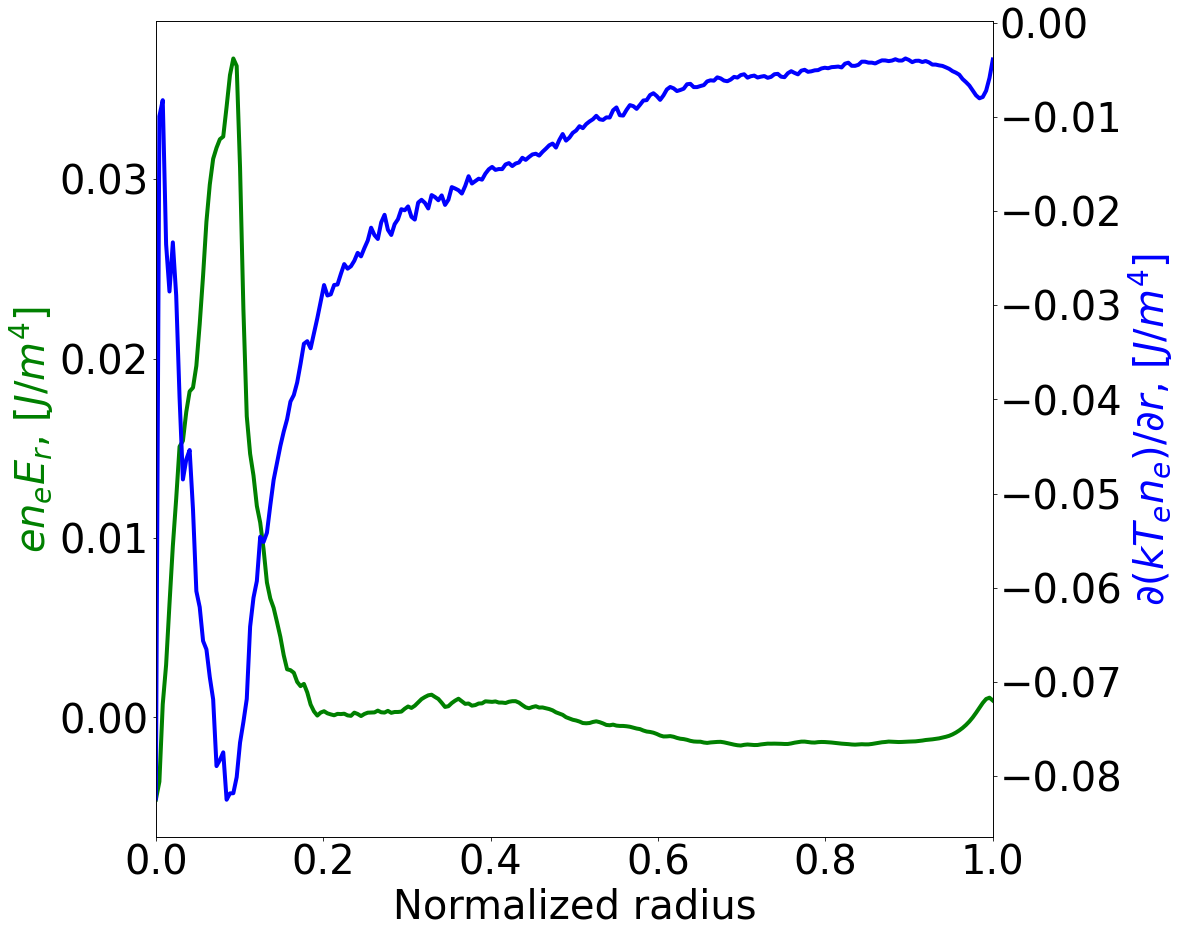}}
\caption{Relative contributions of the electric field and pressure gradient: a) spoke regime,  $B=200$ G; b) spiral arms regime,  $B=150$ G.}
\label{fig:egradp}
\end{figure}

\begin{figure}[H]
\centering
\subcaptionbox{\label{fig:Anommob_B200}}{\includegraphics[width=0.49\linewidth]{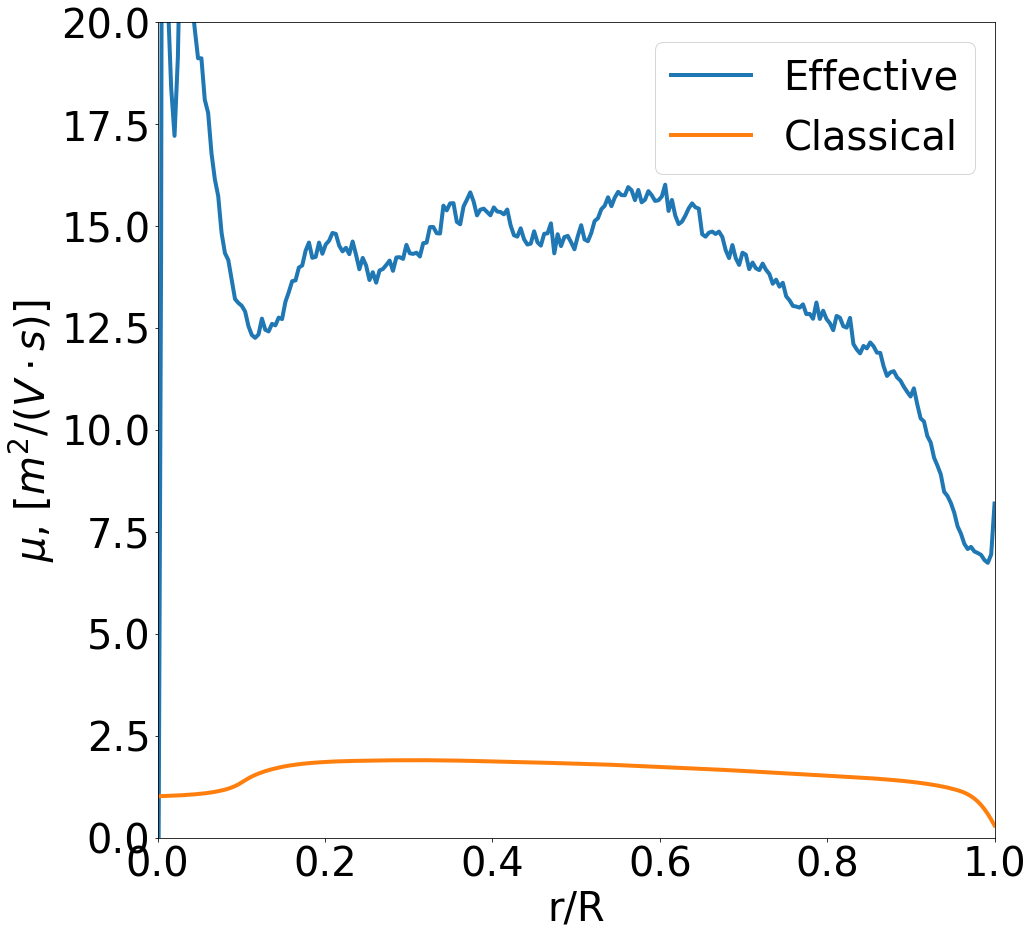}}
\subcaptionbox{\label{fig:Anommob_B150}}{\includegraphics[width=0.47\linewidth]{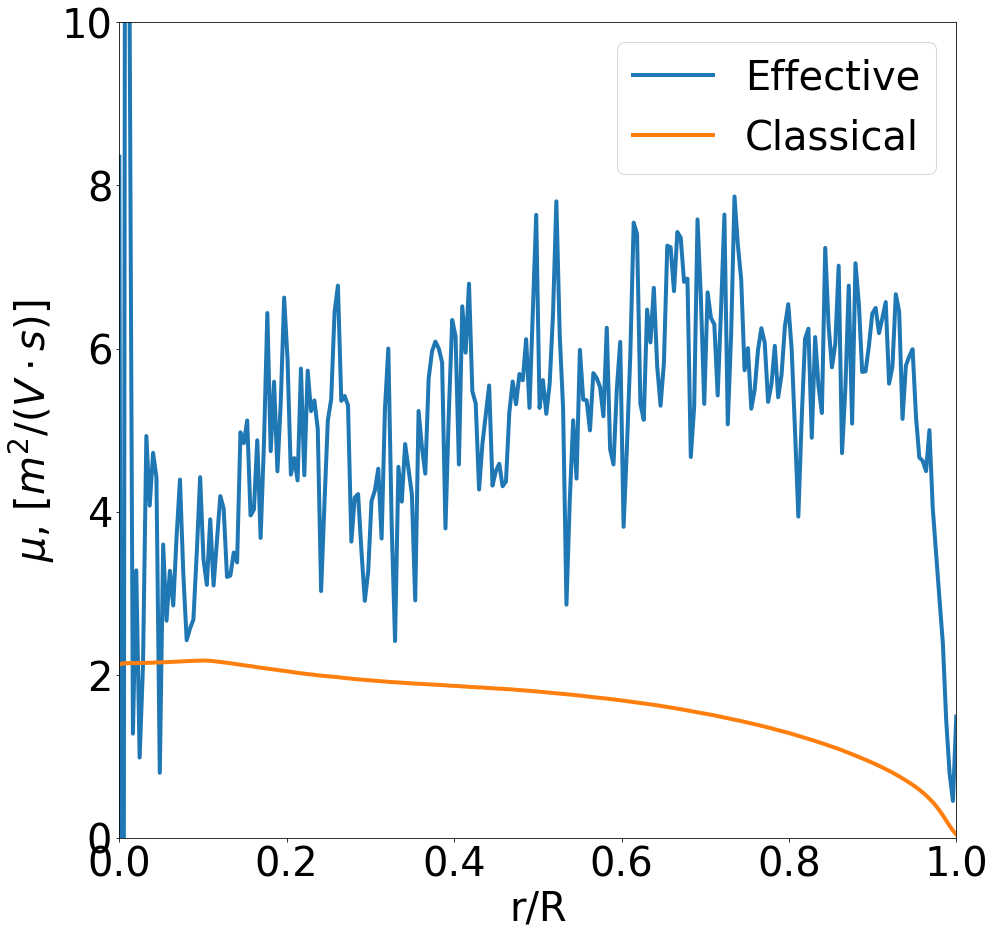}}
\caption{Anomalous mobility: a) spoke regime,  $B=200$ G; b) spiral arms regime, $B=150$ G.}
\label{fig:Anommob}
\end{figure}

In this section  we discuss and compare  the magnitude of the radial current in the m=1 spoke and small scale modes regimes. The radial profiles of the electron and ion current are shown in \Cref{fig:AnomJ}. The spoke regime shows much larger total current.  We characterize the net radial transport as the effective mobility and diffusion by using the representation in the form 
\begin{equation}
 j_e(r) = \mu_{eff}(r)\left[en_e E_r(r) + \frac{\partial}{\partial r} n_e(r)T_{e}(r)\right]. 
\end{equation}
The  radial mobility profiles for $\mu_{eff}$ 
and contributions of the radial electric field (mobility) and pressure gradient (diffusion) are shown in \Cref{fig:Anommob} and \Cref{fig:egradp}. The effective mobility for the spoke regimes is almost order higher that classical values  and for the spiral arms regimes it is several times larger. The relative contributions of the mobility and diffusion are shown in \Cref{fig:egradp}. For the spoke regime, the mobility flux (due to the radial electric field)  is dominant, whereas the diffusion (due to the radial pressure gradient) part is prevailing in the  spiral arms regime.

It is useful to consider the relative contributions of fluctuations and classical transport to the radial current. Neglecting inertia in the electron  momentum equation and  separating the stationary and fluctuation parts in the electric field and density,  $\boldsymbol{E}=\boldsymbol{E_0}+\boldsymbol{\Tilde{E}}$,  $n=n_0 + \Tilde{n}$,  one obtains an equation for the radial electron current:


\begin{equation}
j_r = \frac{\nu e^2 n_0}{(1 + \nu^2/\omega_c^2)m_e \omega^2_c} \left(E_r + \frac{1}{en_0} \frac{\partial p}{\partial r} \right)+
\frac{e^2 \nu}{m_e \omega_c^2} \frac{\langle \Tilde{E_r} \Tilde{n} \rangle}{1 + \nu^2/\omega_c^2} - 
\frac{e^2}{ m_e \omega_c} \frac{\langle \Tilde{n} \Tilde{E_\theta} \rangle}{1 + \nu^2 / \omega_c^2}
\label{Eq:jrfluc}
\end{equation} 

\begin{figure}[H]
\centering
\subcaptionbox{\label{fig:jrtime_B200}}{\includegraphics[width=0.47\linewidth]{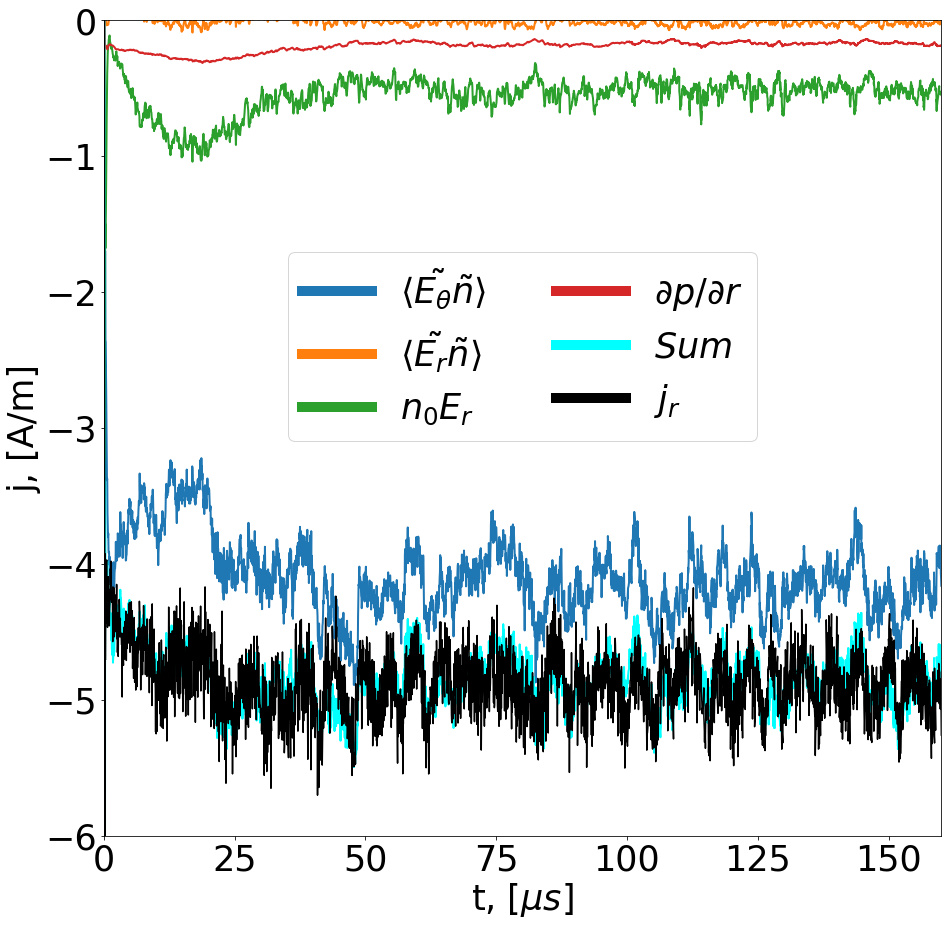}}
\subcaptionbox{\label{fig:jrtime_B150}}{\includegraphics[width=0.49\linewidth]{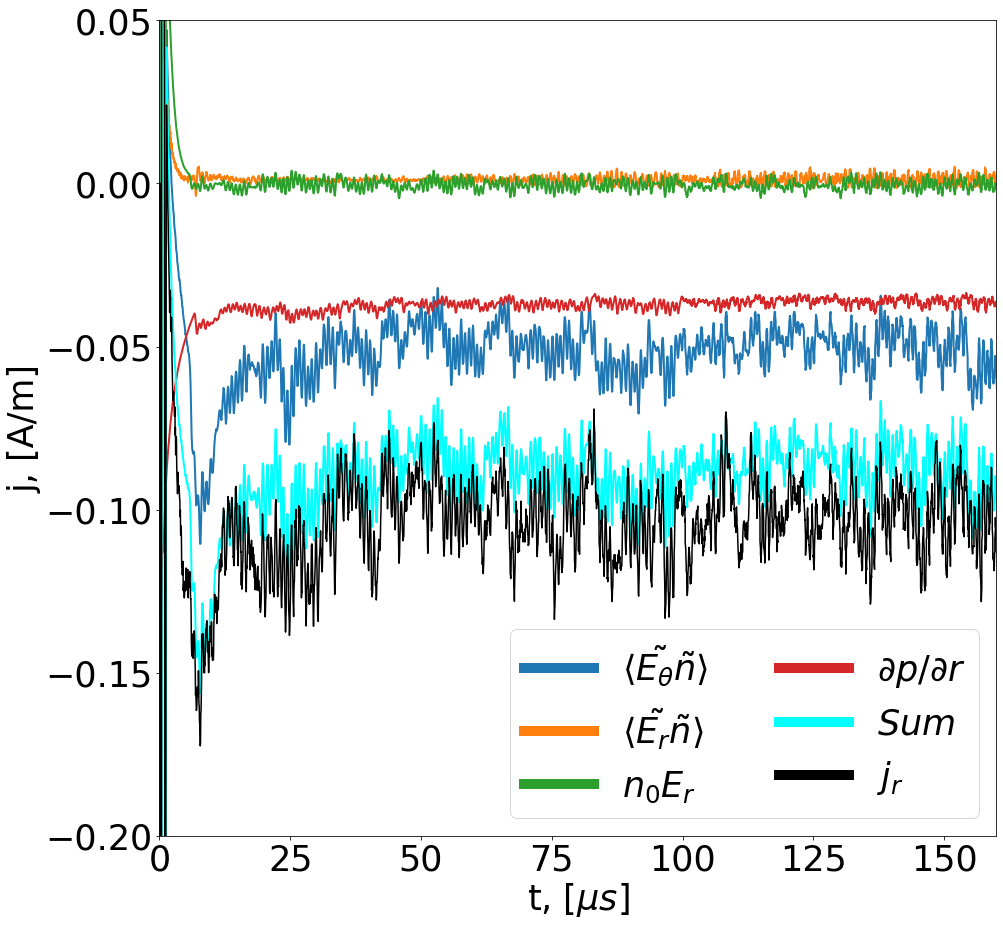}}
\captionsetup{justification=raggedright,singlelinecheck=false}
\caption{Relative contributions of different terms in  \Cref{Eq:jrfluc} terms with time: a) spoke (200 G) and b) spiral arm (150 G) cases.  All terms are averaged over the whole domain. In the case of spiral arms, a moving average was applied to reduce noise.} 
\label{fig:jrtime}
\end{figure}

Two last terms in this equation describe the turbulent transport, while the first two are the classical (collisional) contributions.  
Averaging  over time and azimuthal domain the  radial electron current  can be compared  to the results from the simulations \Cref{fig:jrtime}. For the spoke case, the dominant contribution is due to the  fluctuations,  $~\langle \Tilde{n} \Tilde{E_\theta} \rangle$, \Cref{fig:jrtime_B200} whereas in spiral arm case the contribution of the fluctuations is of the same order as from the collisional diffusion, \Cref{fig:jrtime_B150}.


\section{Summary and discussion}

In this paper, we have studied fluctuations in  the cylindrical Penning discharge self-consistently supported by the ionization from the electron beam. The discharge saturation  to the steady-state was investigated with four different PIC codes: EDIPIC-2D, PEC2PIC, Vsim, and XOOPIC.  It was found that while all codes show qualitatively same behaviour, some quantitative differences in plasma density  (up to the order of 20\%) may occur  due to the differences of  how the particle collisions and relevant cross-sections are implemented.    

 Two different regimes of turbulent structures  and transport were revealed  in the simulations:  the $m=1$ spoke regime with large level of the anomalous transport  and the regime with small scale $m>1$ spiral arm  azimuthal structures.  A characteristic feature of the spiral arm regime is the flat profile of the potential, while in the spoke regime there is a deep potential well confining ions radially. The spiral arm regime is much quieter and has the turbulent transport which is only several times larger than the classical due to collisions. 
 One has to note that the transition between two regimes is continuous and low $m$ (spoke) modes and high $m$ (spiral arm) modes co-exist. In the case of the spoke, most of the power (in the sense of the FFT power spectrum density)  is contained within a few first low-frequency modes (spoke modes $10 \sim 100$ kHz). On the contrary, in the spiral arms regime, the low $m$ modes are weak, and the power is concentrated within a wide range of high-frequency modes (spiral arms $0.4 \sim 1$ MHz).

One of the goals was the investigation of the nature of large scale m=1 spoke activity often observed  in experiments \cite{RaitsesIEPC2015,RodriguezPoP2019}. The comparison of the fluctuations spectra in the spoke regime, obtained  with MUSIC and FFT algorithms, shows reasonable agreement with the theoretical dispersion relation and demonstrate  the existence of the inverse cascade of energy towards the large scale. 
 We have performed parametric studies of the spoke frequency with the magnetic field, simulation box size, and ion species which in general are similar those found in Ref.~\onlinecite{PowisPoP2018}. It is demonstrated that the spoke frequency is mostly determined by the radial electric field and follows the equilibrium ion rotation frequency. The observed dependence on the magnetic field is due to the variation of the radial electric field $E_r$: for larger values of the magnetic field plasma confinement is improved resulting in the  increase of  the $E_r$ and the spoke rotation frequency  according to Equation \ref{Eq:ion_rot}. We have shown that the results are not strongly affected by the difference in the discharge geometry of a similar size. Square,  circular, and  dodecagon geometry show very similar behavior.   
 
 As it was explained above, one of the limitations of the present study is neglect of ion-neutral collisions and ion losses in the axial direction. This is left for future work.

\acknowledgments{This work is supported in part by NSERC Canada, US Air Force Office of
Scientific Research FA9550-21-1-0031 and Plasma Collaborative Research Facility project at Princeton Plasma Physics Laboratory. The computational resources were provided by Compute Canada. The authors would like to thank Tech-X for providing the license for Vsim code.  We acknowledge useful discussions with A. Likhansky. }

\section*{Data availability}
The data that support the findings of this study are available from the corresponding author upon reasonable request.

\bibliographystyle{unsrt}
\bibliography{ref}

\appendix
\section{Simulation Codes}
In this study, we use four PIC codes (EDIPIC, VSim, PEC2PIC, and XOOPIC) to investigate and compare the effect of particle collisions in the Penning discharge supported by the electron  beam. The codes are independent and differ in many detail, but the main differences are in the way the Poisson equation is solved and the MCC-cross section are calculated. A brief summary of simulation techniques used in each code is provided in (\Cref{table:SummaryCh}).

{\bf EDIPIC-2D} is an 2D3V-PIC code developed by D.Sydorenko for simulation of low-temperature plasmas in electrostatic approximation \cite{SydorenkoTh2006}. Trajectories of charged particles are integrated according to leap-frog scheme with the Boris algorithm. The Poisson equation is solved using Generalized minimal residual (GMRes) with tolerance $1\times 10^{-9}$ using  PETSc \cite{PETSc}.
The code uses Monte-Carlo method with  scattering cross sections of electron-neutral and ionization collision from data from LXcat databases. For Argon - elastic and ionization is from Hayashi database, www.lxcat.net, retrieved on May 24, 2021, and excitation from SIGLO database, www.lxcat.net, retrieved on May 24, 2021. For Krypton - Biagi-v7.1 database, www.lxcat.net, retrieved on November 2, 2021. For Nitrogen- BSR-690\_N database, www.lxcat.net, retrieved on November 2, 2021. For Hydrogen- Morgan database, www.lxcat.net, retrieved on November 11, 2021.  It  uses Well Equidistributed Long-period Linear (WELL) algorithm as a pseudo-random number generator \cite{Panneton2006ImprovedLG}.

{\bf VSim} is a proprietary 3D3V-PIC code developed by Tech-X corporation for complex multi physics problem. Here we use it for plasma simulation in  the  electrostatic setting \cite{nieter2004vorpal}. VSim uses a Vorpal computation engine and comes up with the VSim composer that provides a graphical user interface. Particle motion is advanced via the leapfrog and Boris schemes. For solving Poisson equation, the iterative method of GMRes with tolerance $1\times 10^{-8}$ from the Aztec library \cite{HerouxMichaelAllen2004Aug} is used. In VSim, data for cross-sections used in the Monte-Carlo collision model can be considered either by interpolating the  LXcat data set or by specifying the fit function that the cross-section. The electron scattering  from neutral background follows anisotropic Vahedi-Surendra algorithm \cite{vah}. Ionization cross-section for atomic argon is used from Morgan database that is updated in  2015, elastic cross section is taken from NIFS-DATA-72 database presented by M.Hayashi that is updated on 2014, and excitation cross-section is used from A.V.Phelps report, updated in 2010. For Hydrogen, the cross sections are taken from Morgan database updated in 2010. 

{\bf PEC2PIC} is the  {\bf P}arallelized {\bf E}lectrostatic {\bf C}artesian {\bf 2}D {\bf P}article-{\bf I}n-{\bf C}ell 2D3V PIC-MCC solver developed M. Sengupta and R. Ganesh\cite{megh3,megh5}. PEC2PIC is intended for the device simulations of low temperature plasma configurations\cite{meghs,meghsr} and is part of a larger suite of 1-3D Electrostatic PIC codes PECXPIC developed by M. Sengupta, R. Ganesh, et al \cite{meghk,kham2}. PEC2PIC operates on a Cartesian mesh with an iterative Successive Over Relaxation (SOR) Poisson solver\cite{recp}; a combination which gives the flexibility to simulate linear as well as curvilinear shapes of Dirichlet boundaries\cite{meght}. The Gauss-Seidel solving unit of the SOR is Open-ACC accelerated on a GPU using the red-black parallelization scheme\cite{cz,ket}. Newtonian dynamics of particle trajectories are solved using the Lie Operator based Chin's exponential splitting integrator\cite{chin,chin21}. Charge interpolation from particle to mesh nodes, and electric field interpolation in the opposite direction are achieved via the first order Cloud-in-Cell (CIC) scheme\cite{bird}. The Particle-push and the CIC are parallelized on a CPU node using Open-MP. An MCC routine based on Vahedi et al.'s algorithms\cite{vah} with an improved electron anisotropic scattering equation \cite{okh} simulates the collisional interaction of plasma with background neutrals. The code's language is Fortran. 

{\bf XOOPIC} (X-windows Object-Oriented Particle In-Cell code) is an open-source 2D3V cartesian (x-y \& r-z) software developed by the Plasma Theory and Simulation Group (PTSG) \cite{XOOPIC}. In this work we use the electrostatic serial version of XOOPIC for non-relativistic equations of motion of  charged particles using  Boris advance technique. Poisson equation is solved by iterative method of Dynamic Alternating Direction Implicit (DADI) with tolerance $1\times 10^{-3}$. The required cross-sections for Monte-Carlo collision are estimated from continuous regression-based functions.

\begin{table}[H]
\caption{Main features of four PIC codes employed in the benchmarking study.}
\scalebox{0.9}{
\scriptsize{
\begin{tabular}{|p{3.5cm}|p{3.5cm}|p{3.5cm}|p{3.5cm}| p{3.5cm}|}
\hline\hline 
Code & EDIPIC & VSim & PEC2PIC & XOOPIC  \\ \hline\hline
\rowcolor{Gray}
\multicolumn{5}{|c|}{\textbf{Algorithm}} \\\hline\hline 
Particle-Mesh assignment & & NA & First order Cloud-in-Cell &\\\hline
Poission Solver& KSP GMRes (PETSc) &{\bf Iterative:} Generalized minimal residual (GMRes) & {\bf Iterative:} Successive Over Relaxation (SOR) &  {\bf Iterative:} Dynamic Alternating Direction Implicit (DADI)  \\ \hline
Push Solver & {\bf Leap Frog:} Boris Advancement & {\bf Leap Frog:} Boris Advancement & {\bf Lie Operator formalism:} Chin's Exponential Splitting & {\bf Leap Frog:} Boris Advancement \\ \hline
MCC-Cross section & {\bf Interpolated} from Lxcat data set& {\bf Interpolated} from {\bf discrete} data set (Lxcat) & {\bf Estimated} from {\bf continuous Regression-based} functions & {\bf Estimated} from {\bf continuous Regression-based} functions\\  \hline
Electron Scattering &  & Anisotropic & Anisotropic & \\ \hline \hline

\rowcolor{Gray}
\multicolumn{5}{|c|}{\textbf{Hardware Acceleration}} \\\hline\hline 
Architecture& CPU & CPU & CPU-GPU & CPU         \\ \hline
Parallelization& MPI & MPI & Open-MP $\&$ Open-ACC &     NA      \\ \hline
Decomposition& Domain & Domain& NA &  NA         \\ \hline
Floating-Point Precision& Double & Double & Double & Double           \\ \hline
Language& Fortran 90 & C++ & Fortran &  C++         \\ \hline\hline
\end{tabular}}}
\label{table:SummaryCh}
\end{table}
\section{The effects of the geometry of the simulation region.}
\begin{figure}[htp]
\centering
\includegraphics[width=0.5\textwidth]{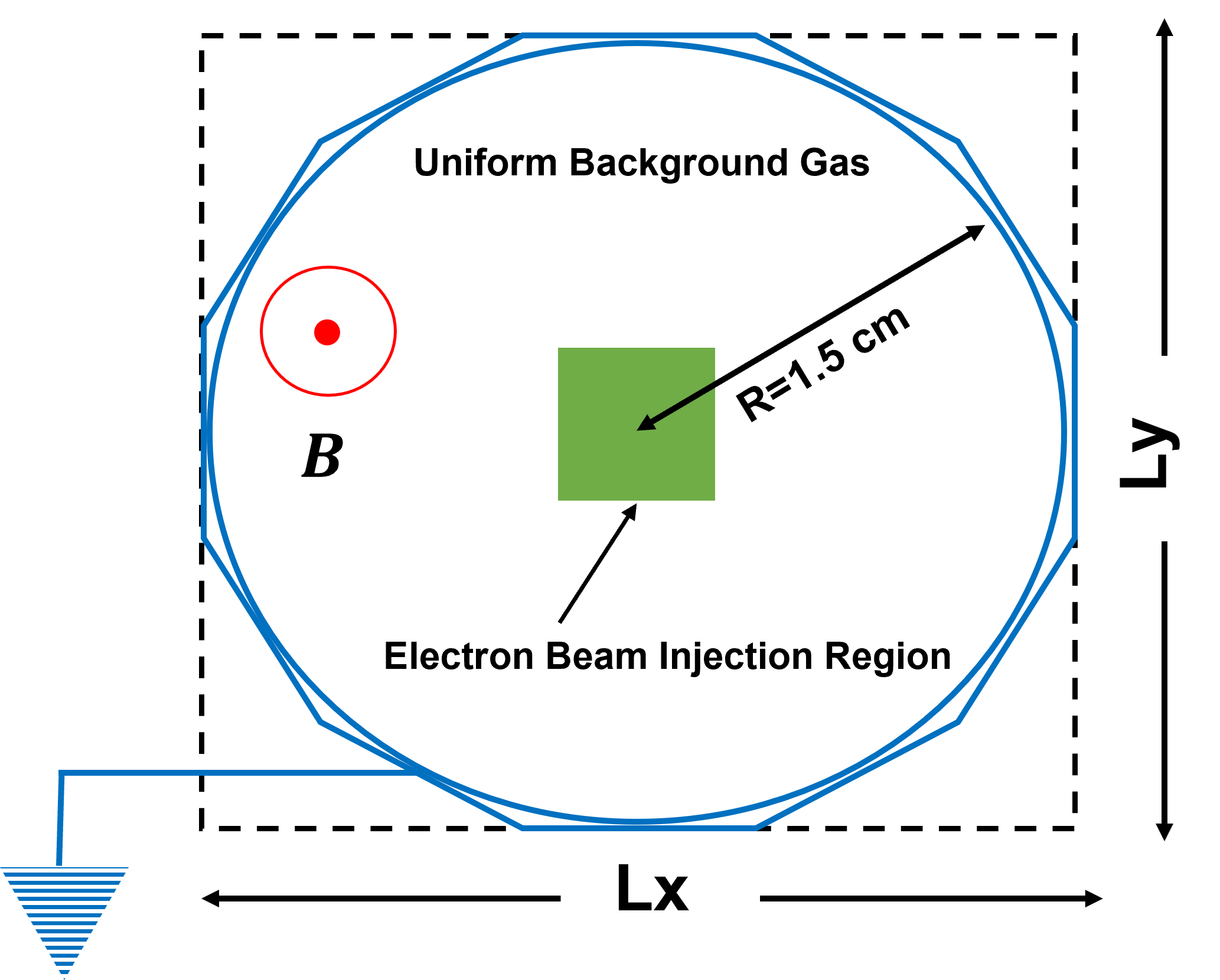}
\captionsetup{justification=raggedright,singlelinecheck=false}
\caption{The end view of the circular cross section used in VSim and PEC2PIC codes and dodecagon cross-section that is used in XOOPIC of the modeled discharge.}
\label{circularDodecagonB}
\end{figure}

Our base case simulations were performed in the square box geometry, while both rectangular geometry and circular geometry are used  in magnetically enhanced  $\mathbf{E}\cross \mathbf{B}$ discharges \cite{Fubiani}. Therefore it is of interest to investigate the differences that occur for different shapes  of the boundary of the simulation region. On this subject we have performed simulations of the case only ionizing collision and ionization plus non-ionizing electron neutral collisions for Argon and Hydrogen with the circular cross section implemented in VSim and PEC2PIC codes. In XOOPIC, we have used the dodecagon boundary. Both circular and dodecagon boundaries are located inside the square box of simulation as shown in \Cref{circularDodecagonB} with a uniform Cartesian mesh grid. The diameter of circular is equal to the sides of the square configuration (\Cref{fig:SimulationSetUp}). The cell size and time step are identical with corresponding square boundary simulation.  

\begin{figure}[htp]
\centering
\subfloat{\label{}\includegraphics[width=0.5\linewidth]{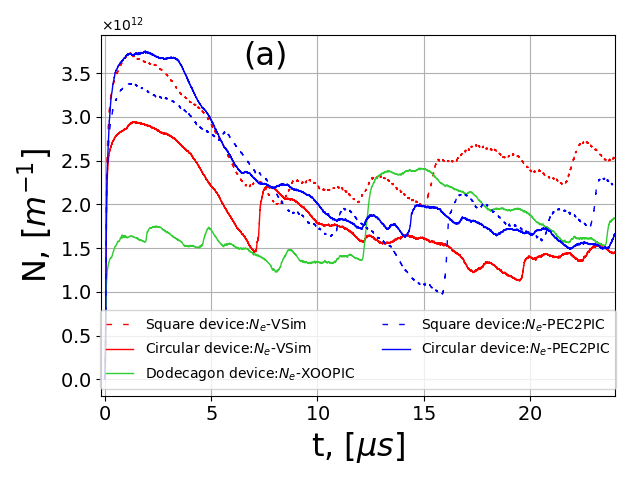}}
\subfloat{\label{}\includegraphics[width=0.5\linewidth]{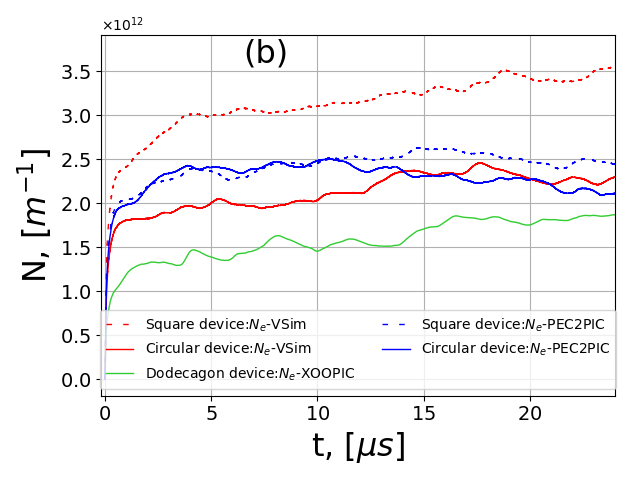}}
\captionsetup{justification=raggedright,singlelinecheck=false}
\caption{Temporal evolution of electrons inventory of Argon gas in circular and square simulation geometry. a) case with only ionization, b) case with ionization plus non-ionizing electron neutral collisions. Dashed lines indicate electrons evolution in square geometry and solid lines show evolution of electrons in circular-like geometry.}
\label{BenchCir}
\end{figure}

\begin{figure}[htp]
\centering
\subfloat{\label{}\includegraphics[width=0.5\linewidth]{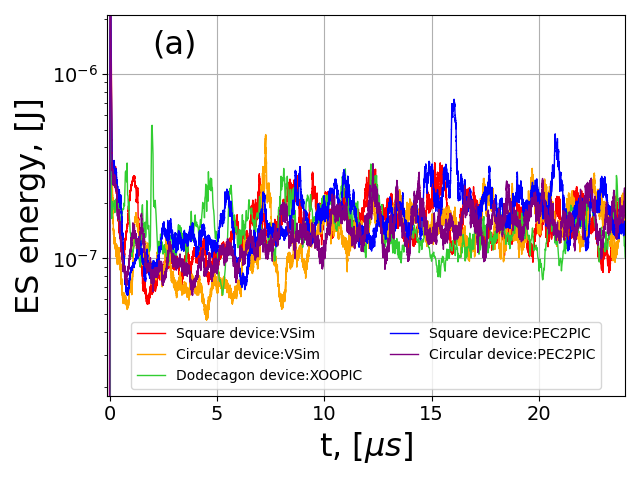}}
\subfloat{\label{}\includegraphics[width=0.5\linewidth]{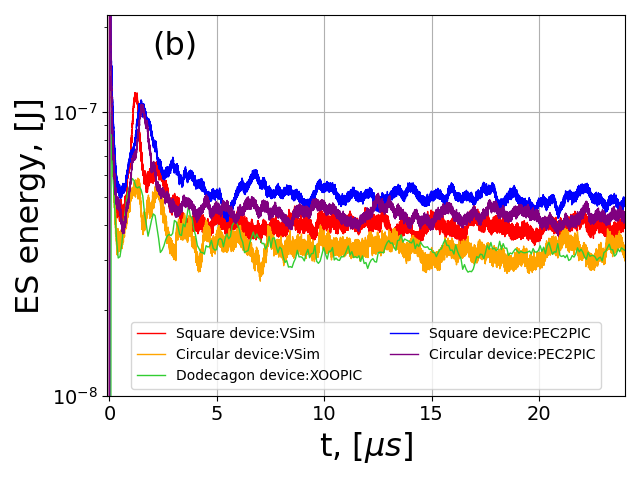}}
\captionsetup{justification=raggedright,singlelinecheck=false}
\caption{Temporal evolution of electrostatic energy of Argon gas in circular and square simulation geometry. a) case with only ionization, b) case with ionization plus non-ionizing electron neutral collisions.}
\label{BenchCirESE}
\end{figure}

\Cref{BenchCir}a shows the evolution of electron inventory of Argon discharge simulation with only ionization, in circular and dodecagon devices in comparison with square device. It can be seen the number of particles in saturation level of circular boundary (solid line) is less than square boundary (dashed line with similar color). \Cref{BenchCir}b which is related to the evolution of electrons in Argon discharge simulation with ionization plus non-ionizing electron neutral collisions, shows similar result that the number of particles in circular boundary are less than the corresponding square boundary. The result of simulations these two cases with dodecagon geometry of XOOPIC code represents good agreement with circular geometry results of VSim and PEC2PIC codes. Figures \ref{BenchCirESE}a and b show that the ES energy does not differ for square and circular boundaries at saturation stage of corresponding simulations.

\begin{figure}[htp]
\centering
\includegraphics[width=0.9\linewidth]{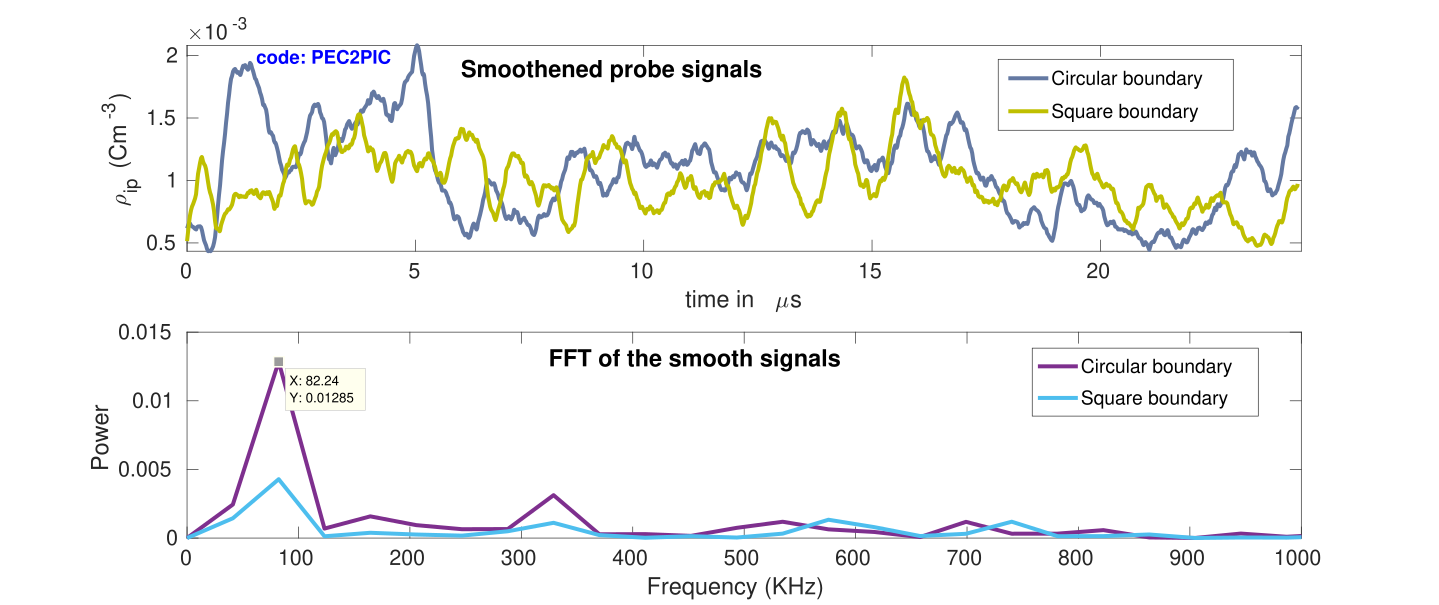}
\captionsetup{justification=raggedright,singlelinecheck=false}
\caption{Comparison of the prob signals for the ion charge density,  $\rho_{ip}$, and power spectrum between square and circle devices for Argon, showing the effect of boundary on spoke frequency.}
\label{ProbeSignalArgon}
\end{figure}
\Cref{ProbeSignalArgon} compares rotation frequency of the spoke in PEC2PIC simulations of the Argon discharge (with ionizing plus non-ionizing electron-neutral interactions) in square and circular devices. The method used for measuring the $m=1$ frequency is the same as in \Cref{fig:PEC2PICProbAnalysis} (i.e. density probe signal analysis). \Cref{ProbeSignalArgon}-top shows the smoothened ion density signal of the probe for the two devices while \Cref{ProbeSignalArgon}-bottom has the corresponding FFT analysis. We get nearly equal spoke rotation frequencies for the devices, indicating the cross-sectional boundary shape has little impact on the steady state spoke's frequency.

\begin{figure}[htp]
\centering
\subfloat{\label{BenchCirNumNoInoH}\includegraphics[width=0.5\linewidth]{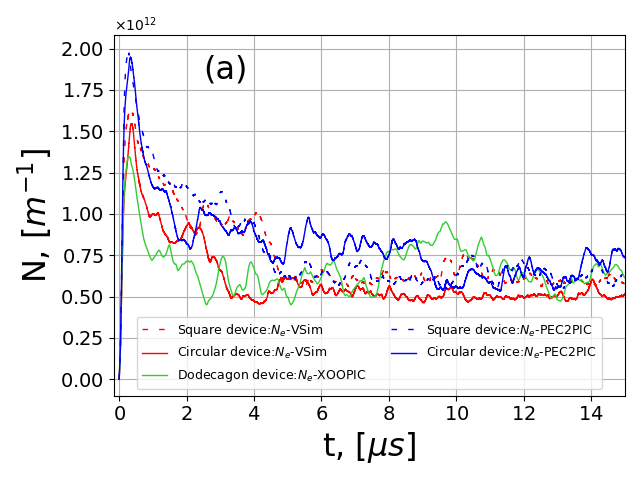}}
\subfloat{\label{BenchCirESENoInoH}\includegraphics[width=0.5\linewidth]{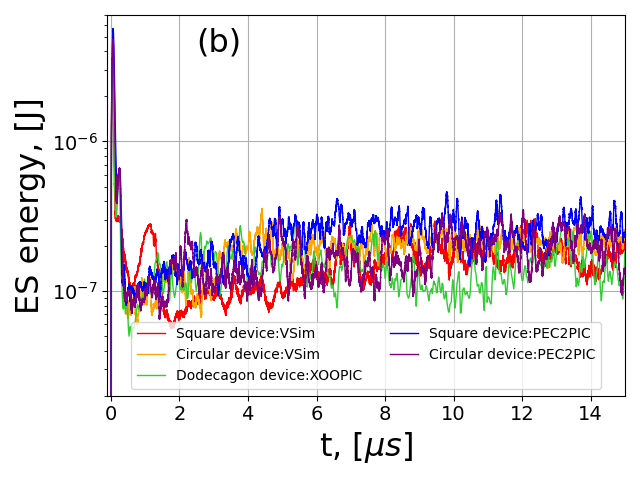}}
\captionsetup{justification=raggedright,singlelinecheck=false}
\caption{Temporal evolution of a) electrons inventory, b) electrostatic energy of Hydrogen gas for simulation with ionization plus non-ionizing electron neutral collision in circular and square simulation geometry.}
\label{BenchCirNoInoH}
\end{figure}
\begin{figure}[htp]
\centering
\includegraphics[width=0.9\linewidth]{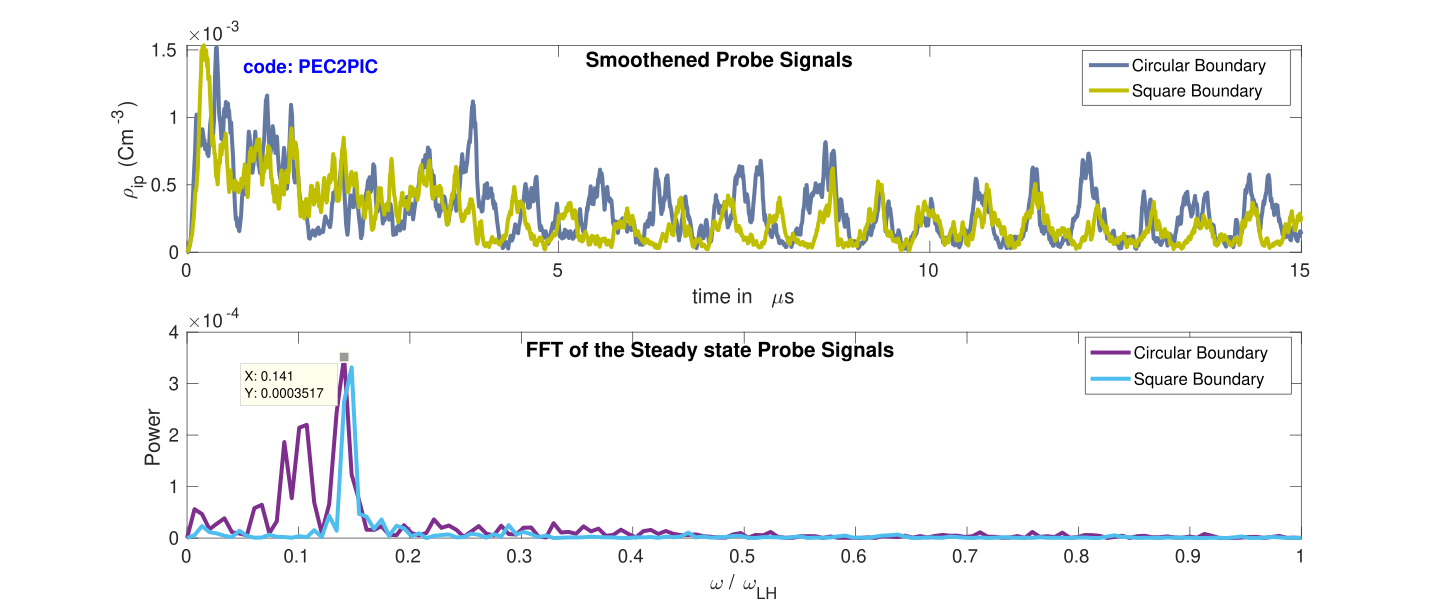}
\captionsetup{justification=raggedright,singlelinecheck=false}
\caption{Comparison of the prob signals for the ion charge density,  $\rho_{ip}$, and power spectrum between square and circle devices for Hydrogen, showing the effect of boundary on spoke frequency.}
\label{ProbeSignalH}
\end{figure}

Applying Hydrogen gas in simulation case with ionization plus non-ionizing electron neutral collisions, instead of Argon gas, reveals that the main features of simulation results, including particle and ES energy evolution are not different for square and circular geometries (see \Cref{BenchCirNoInoH}). \Cref{ProbeSignalH} compares rotation frequency of the spoke in PEC2PIC simulations of the Hydrogen discharge (with ionizing plus non-ionizing electron-neutral interactions) in equal sized square and circular devices. \Cref{ProbeSignalH}-top has ion density signals at the probe and \Cref{ProbeSignalH}-bottom has the corresponding FFTs. Unlike Argon, the Hydrogen discharge shows significant differences in frequency spectrum for the two device shapes. While the $m=1$ frequency for the two configurations are close , about $1.38\;\text{MHz}$ the circular device has additional peaks at sub $m=1$ frequencies. The source of these additional peaks needs further investigation to understand.


\nocite{*}
\end{document}